\documentclass[fleqn,usenatbib]{mnras}

\usepackage{graphicx}
\usepackage{longtable}
\usepackage{url}
\usepackage{epstopdf}
\usepackage{color}
\usepackage{bm}
\usepackage{amssymb}
\usepackage{amsmath}
\usepackage{times}
\usepackage{nccmath}

\usepackage{hyperref}
\hypersetup{
   colorlinks=true,       
    linkcolor=red,          
    citecolor=blue,        
    filecolor=magenta,      
    urlcolor=cyan           
}

\newcommand{\etal}{{\em et al. }}
\newcommand{\ijmpa}{International Journal of Modern Physics A}

\DeclareTextSymbol{\degre}{OT1}{23}
\newcommand{\cig}{\textsc{cigale}}
\newcommand{\cigbig}{\textsc{CIGALE}}

\newcommand{\lephare}{\textsc{lephare}}
\newcommand{\getcv}{\textsc{getcv}}

\newcommand{\lsi}{$L-{\rm SFR}$}
\newcommand{\zph}{$z_{ph}$}

\newcommand{\lgmstar}{$\log_{10}(M_* / M_\odot)$}
\newcommand{\mstar}{$M_*$}

\newcommand{\mic}{$\mu$m}

\newcommand{\her}{{\it Herschel}}

\newcommand{\sig}{$1\sigma$} 
\newcommand{\cosmo}{\textsc{CosmoMC}}
\newcommand{\sigv}{$\sigma_{\rm V}$}

\newcommand{\errtot}{$\sigma_{\rm tot}$}
\newcommand{\sigms}{${\sigma_{MS}}$}
\newcommand{\sigmsmean}{${\bar\sigma_{MS}}$}
\newcommand{\whit}{W12}

\newcommand{\ur}{{\it u-r}}
\newcommand{\mph}{Mpc\ $h^{-1}$}

\newcommand{\sfr}{SFR }

\title[The evolving (s)SFR-\mstar\ relation in the VIDEO Survey]{The evolving relation between star-formation rate and stellar mass in the  VIDEO Survey since $z=3$}

\author[Johnston R, \etal]{
Russell Johnston,$^{1,5}$\thanks{rwi.johnston@gmail.com}
Mattia Vaccari,$^{1,6}$
Matt Jarvis,$^{2,1}$
Mathew Smith,$^{1,4}$
Elodie Giovannoli,$^{1}$
\newauthor Boris  H\"au{\ss}ler$^{2,3}$
and Matthew Prescott$^1$
\\
$^1$Physics Department, University of the Western Cape, Cape Town 7535, South Africa\\
$^2$Astrophysics, University of Oxford, Keble Road, Oxford, OX1 3RH, UK\\
$^3$Centre for Astrophysics, Science \& Technology Research Institute, University of Hertfordshire, Hatfield, Herts AL10 9AB, UK\\
$^4$School of Physics and Astronomy, University of Southampton, Southampton SO17 1BJ\\
$^5$South African Astronomical Observatory, P.O. Box 9, Observatory 7935, South Africa\\
$^6$INAF - Istituto di Radioastronomia, via Gobetti 101, 40129 Bologna, Italy
}

\date{Accepted XXX. Received YYY; in original form ZZZ}

\pubyear{2015}

\begin{document}
\label{firstpage}
\pagerange{\pageref{firstpage}--\pageref{lastpage}} \pubyear{2011}
\maketitle

\begin{abstract}
We investigate the star-formation rate (SFR) and stellar mass, \mstar\ relation  of a  star-forming  (SF) galaxy sample  in the XMM-LSS field to $z\sim 3.0$ using the near-infrared data from the  VISTA Deep Extragalactic Observations (VIDEO) survey. Combining VIDEO with  broad-band photometry,  we use the SED fitting algorithm \cig\ to derive SFRs and \mstar\ and have adapted it to account for the full photometric redshift PDF uncertainty.  Applying a SF selection using the D4000 index, we find evidence for strong evolution in the normalisation of the SFR-\mstar\ relation out to $z\sim 3$ and a roughly constant slope of  (SFR $\propto M_*^{\alpha}$) $\alpha=0.69\pm0.02$ to $z\sim 1.7$. We find this increases close to unity toward $z\sim2.65$. Alternatively, if we apply a colour selection, we find a distinct  turnover in the SFR-\mstar\ relation between $0.7\lesssim z\lesssim2.0$ at the high mass end, and suggest that this is due to an increased contamination from passive galaxies. We  find evolution of the specific SFR  $\propto(1+z)^{2.60}$ at \lgmstar~$\sim$~10.5,  out  to  $z\lesssim2.4$ with an observed flattening beyond $z\sim$ 2 with increased stellar mass.  Comparing to a range of simulations we find the analytical scaling relation approaches, that invoke an equilibrium model, a good fit to our  data, suggesting that a continual smooth accretion regulated by continual outflows may be a key driver in the overall growth of SFGs.\\
\end{abstract}

\section{Introduction}
Quantifying the interplay between star formation and stellar mass (\mstar) is a crucial component to understanding the build up of galaxies over cosmic time.  Whilst early works established the decline in SFR as a function of redshift \citep[e.g.][]{madau95,Lilly1996ApJ...460L...1L,Hopkins2006ApJ...651..142H}, the recent acquisition of data from  broad band photometry coupled with access to deep and wide galaxy surveys, led to \cite{Noeske2007ApJ...660L..43N}  demonstrating a  strong correlation between star-formation rate (SFR) and \mstar\ out to $z\sim1$, which they termed the star forming `Main Sequence' (SF-MS). This SF-MS was modelled by a simple power-law of the form 
SFR=$\beta M_*^\alpha$ where strong evolution in the normalisation results in a rapid decline in SFR from $z\sim1$ to $z\sim0.3$ (at a fixed stellar mass), with a slope $\alpha\sim0.7$ at $z\sim0.5$. 
 Complementary  studies at this time by \cite{Elbaz2007AA...468...33E} and \cite{daddi07} found  a similar trend with slopes close to unity at $z\sim1$ and 2 respectively.   
 Subsequent work showed an  anti-correlation between the specific SFR and stellar mass  out to $z\sim 2$ which has been termed {\it downsizing} \citep{Cowie1996AJ....112..839C} where the more  massive galaxies form at higher redshift  
 \citep[see e.g.][]{Noeske:2007ApJ...660L..47N, Damen:2009ApJ...690..937D,Rodighiero2010}.
\begin{figure}
\centering
 \includegraphics[width=0.49\textwidth]{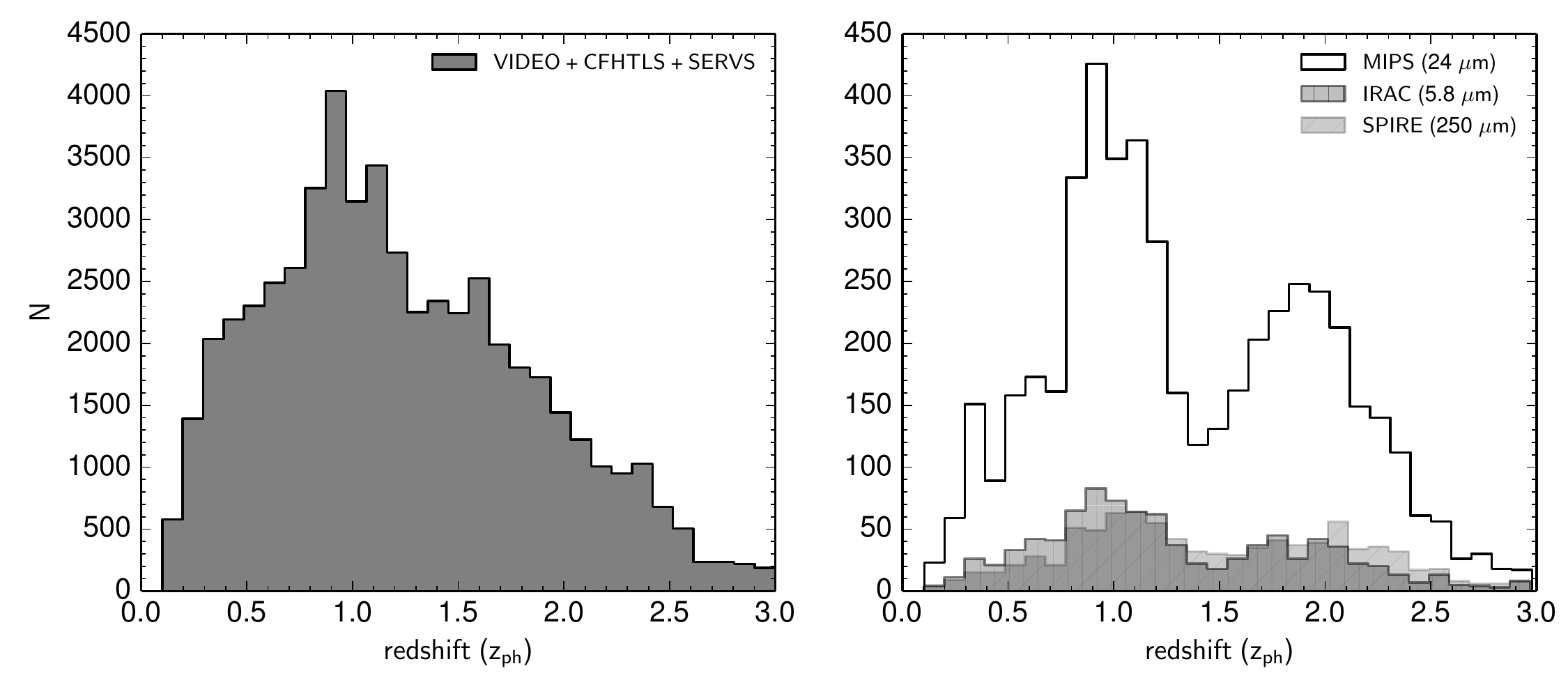}
     \caption{\small  {\it Left:} The VIDEO redshift distribution from our SERVS-matched sample of 52,812 galaxies. {\it Right:} The contribution to the redshift distribution by sources detected by MIPS 24 (4881 galaxies),  IRAC  5.8 (909 galaxies) and SPIRE 250\,\mic\ (901 galaxies).  }
    \label{fig:dndz}    
\end{figure}

Since then, there have been many investigations of this relationship, using both observations and simulations, with the aim of shedding light on how it connects with the underlying physical processes governing galaxy evolution. Whilst the general trends in the SF-MS from observations seem to agree,  the way in which we obtain SFR measurements, coupled with the choice of initial mass function (IMF), the particular star-forming galaxy (SFG) selection criterion, the modelling of star formation histories (SFH) and stellar population synthesis models (SPS), dust attenuation, extinction, photometric redshifts, metallicities,  adopted cosmology and incompleteness, can all introduce bias and calibration issues for modelling the SFR-$M_*$  relation accurately \citep[e.g.][]{Salimbeni2009AIPC.1111..207S,Conroy2009ApJ...699..486C,Magdis2010MNRAS.401.1521M,Walcher2011ApSS.331....1W,Pforr2012MNRAS.422.3285P,Conroy2013ARAA..51..393C,Buat:2014AA...561A..39B,Rodighiero2014MNRAS.443...19R,Speagle2014ApJS..214...15S}.    

Typical SFR indicators include the  UV-wavelength range  \citep[e.g.][]{Erb2006ApJ...647..128E,Magdis2010MNRAS.401.1521M,Rodighiero2011ApJ...739L..40R,leeKS11,Lee2012ApJ...752...66L} since the emission is dominated by young massive short-lived stars. The UV, however, is heavily affected by extinction and so combining this indicator with IR measurements helps to correct for   dust attenuation in the UV which is then re-emitted in the IR and FIR  \citep[e.g.][]{Santini2009AA...504..751S,Salmi2012ApJ...754L..14S,Whitaker2012,Reddy2012ApJ...754...25R,Smith2012}, or using the IR alone as in e.g. \citet{Oliver2010MNRAS.405.2279O,Elbaz2011AA...533A.119E}.  Nebula emission lines,  e.g. H$\alpha$, O{\sc{[ii]}} and O{\sc{[iii]}}  have also been used \citep[e.g.][]{Chen2009MNRAS.393..406C,Shim2011ApJ...738...69S,Zahid2012ApJ...757...54Z,Suzuki2015arXiv150502410S}, as well as from  radio continuum emission and stacking techniques \citep[e.g.][]{Dunne:2009MNRAS.394....3D,Pannella2009ApJ...698L.116P,Karim2011ApJ...730...61K,Zwart2014MNRAS.439.1459Z}.  
Recently, SED modelling algorithms using so-called energy balance methods that exploit the conservation of energy between the stellar light absorbed by dust that is then re-emitted in the mid-IR/far-IR/submm  \citep{burgarella05b,noll09b,dC08}, have also been used to obtain estimates of the total SFR as well as stellar masses and other physical properties.

All of these approaches have advantages and disadvantages. However,  a key issue is the initial selection of SFGs \citep[e.g.][]{Ilbert2010ApJ...709..644I, Karim2011ApJ...730...61K}. Recent studies  by \cite{Rodighiero2014MNRAS.443...19R} and \cite{Speagle2014ApJS..214...15S} demonstrate how different colour selection of SFGs leads to large variations in determining the slope  of the SF-MS throughout the literature, making direct comparisons challenging.

Nevertheless, some of the most recent work presents evidence for a more complex evolving MS beyond a standard power-law. For example, \cite{Heinis:2014MNRAS.437.1268H,Magnelli2014AA...561A..86M,Schreiber2015AA...575A..74S} and 
\cite{Whitaker2014ApJ...795..104W} present results that indicate a MS that flattens off toward the high end of the mass function, favouring a MS modelled by e.g. a broken power-law. In particular, \cite{Whitaker2014ApJ...795..104W}, using  data from the 3D-HST photometric catalogues and a stacking analysis of  MIPS 24\mic\ probed the low mass SF-MS down to \lgmstar = 8.4 and 9.2, at z=0.5 and 2.5 respectively. From this analysis they parameterised the data with a two-power law model, and found evidence for  strong redshift evolution for galaxies with masses \lgmstar$>10.2$. However, they found no evolution for galaxies below this mass limit, suggesting that high-mass SFGs may follow a different evolutionary path than lower mass galaxies. This trend at the high-mass end has also been reported by \cite{Tasca2014arXiv1411.5687T} using data from the VIMOS Ultra-Deep Survey (VUDS) at $z\sim1.5$.

\begin{figure}
\centering
 \includegraphics[width=0.49\textwidth]{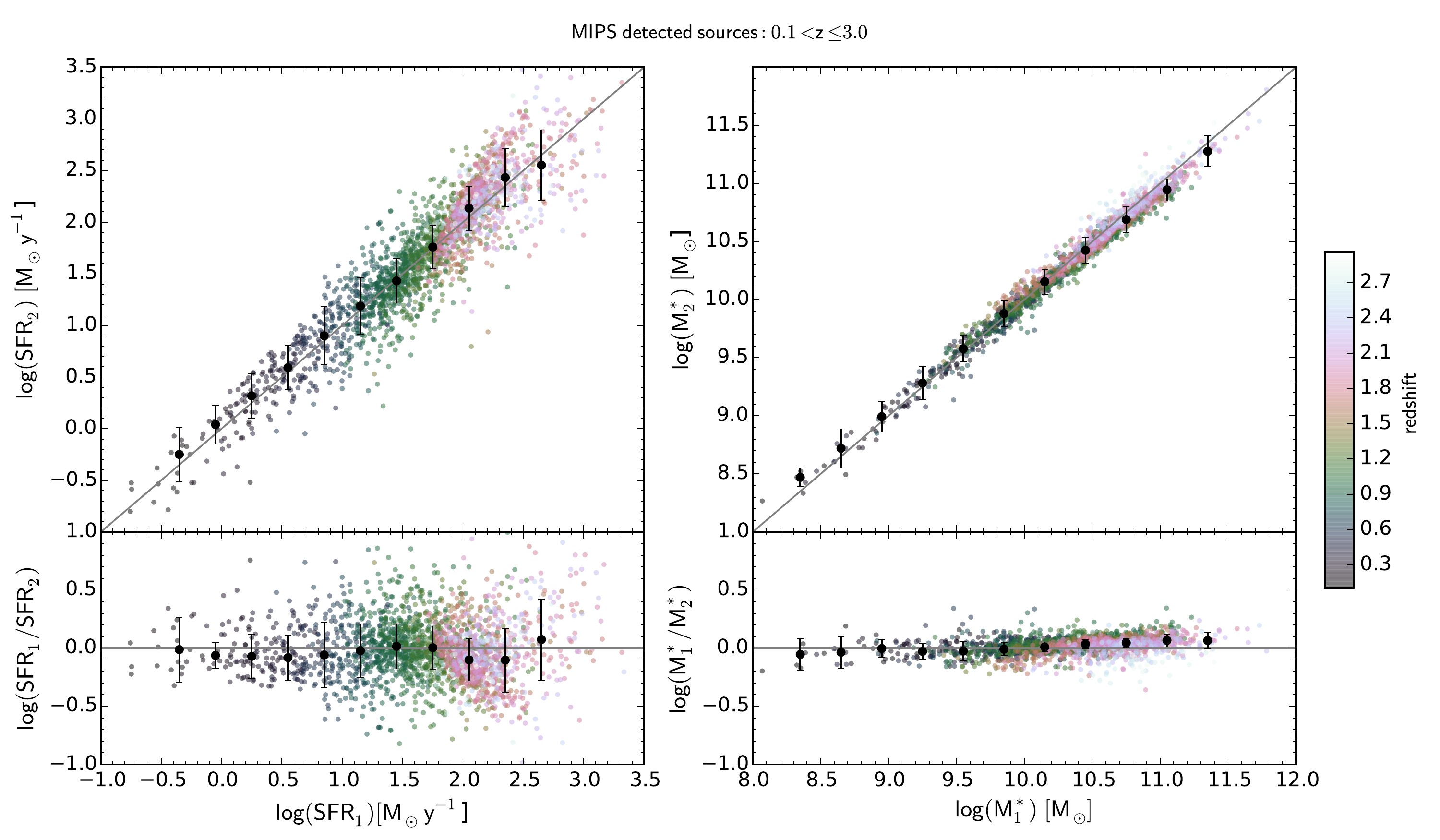}
     \caption{\small Comparison of  the outputs from \cig\ using the full wavelength coverage available and a restricted coverage.  The left-hand panels show a comparison of SFR estimates from \cig\  where SFR$_1$ is the resulting SFR using only the input photometry from VIDEO+CFHT+SERVS. The label SFR$_2$ is for the same set of objects when we include additional  filter information from MIPS+IRAC+\her. The right-hand panel shows the same comparison for the stellar masses $M_1^*$ and $M_2^*$. The black points in each panel are the medians with the  $\sigma_{MAD}$ errors. The grey solid lines on the top left and right panels indicates the one-to-one relation. The colour scale indicates increasing redshift from blue to red which covers $0.1\lesssim z \lesssim 3.0$. The bottom two panels show the $\log_{10}$ of the ratio between SFR$_1$ and SFR$_2$ ({\em left}) and the stellar masses ({\em right}). }
    \label{fig:mips}    
\end{figure} 
In parallel, over the last decade there have been significant advances in simulations providing insight into  galaxy formation through the modelling of physical and environmental processes. The most popular approaches that  have emerged are the semi-analytical models (SAMs) and hydrodynamical simulations. SAMs use an analytical approach, requiring up to  $\sim$50 parameters to characterise the gas dynamics within and around halos \citep[e.g.][]{Bower2006MNRAS.370..645B,Kitzbichler:2007MNRAS.376....2K,Bower2008MNRAS.390.1399B,Guo2008MNRAS.384....2G}.  Whilst they have been largely successful at reproducing observed SFRs of local populations ($z\lesssim0.4$),  the work by e.g.  \cite{Elbaz2007AA...468...33E,daddi07,Santini2009AA...504..751S} and \cite{Damen2009ApJ...705..617D} have  shown they consistently  under predict the star formation rate by a factor of 2-5 out to $z\sim2$.

With the advances in computational power,  hydrodynamical simulations have grown in popularity. This approach attempts to model the physical dynamics of cosmic gas in a self consistent way by including the modelling of  star formation processes, feedback from stellar winds, supernovae, and black holes \citep[e.g.][]{Dav2011MNRAS.415...11D,Dave2013MNRAS.434.2645D, Dubois2014MNRAS.444.1453D, Vogelsberger2014Natur.509..177V,Sparre:2014arXiv1409.0009S}. However, while there may be broad agreement with overall trends of the SFR-\mstar\ relation with observations, they too exhibit a  consistent under prediction in the overall normalisation across most redshifts \citep[e.g.][]{daddi07,Tasca2014arXiv1411.5687T}.

\begin{figure}
\centering
 \includegraphics[width=0.48\textwidth]{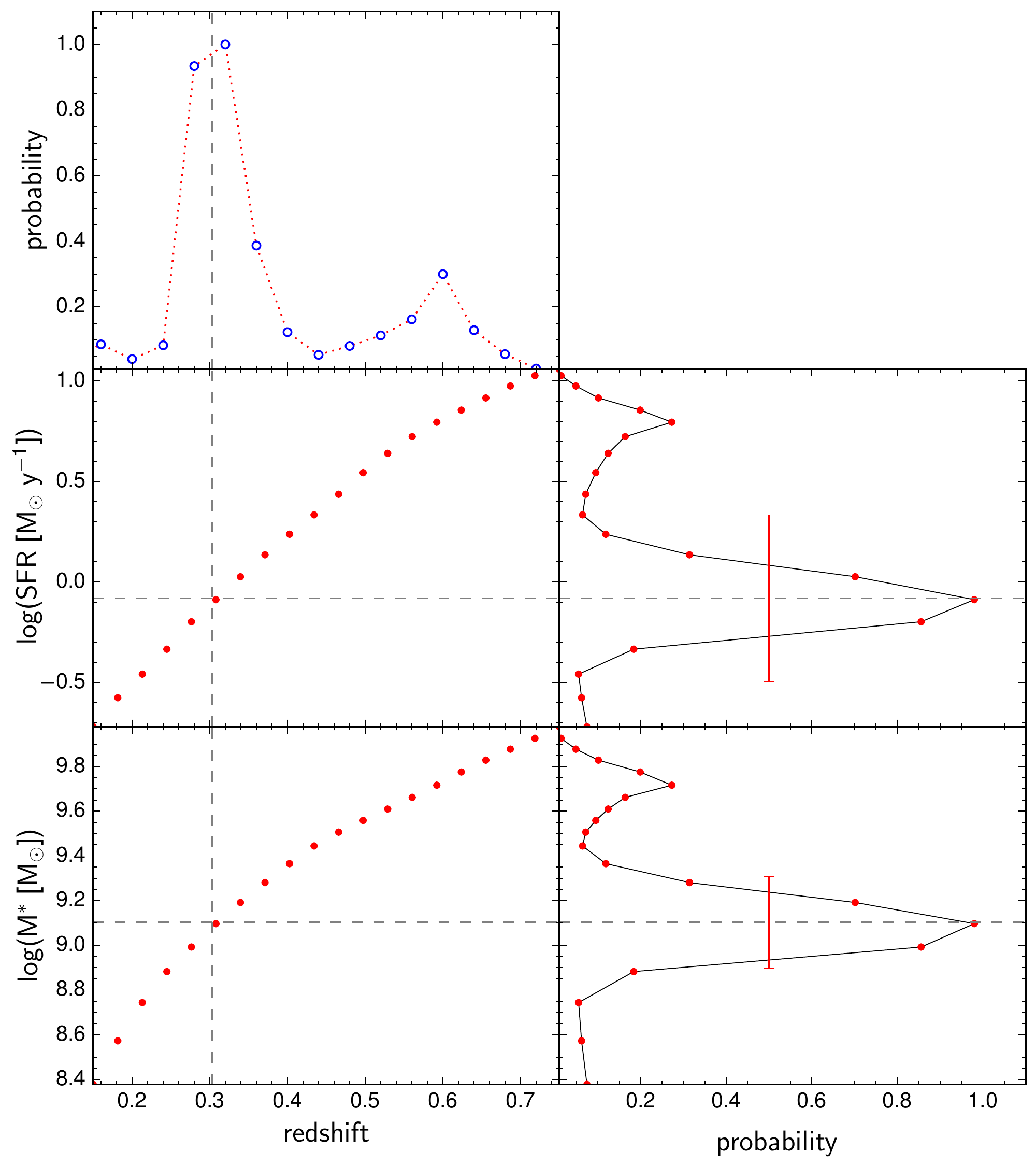}
     \caption{\small Example illustrating how we propagate photometric redshift PDF estimates through \cig. The top panel shows  a typical redshift PDF outputted by \lephare\ and the blue open circles represent the gridded redshift steps.  The best-fit redshift for this object is shown by the vertical dashed line.  To estimate the resulting PDF for the SFR and \mstar, we run \cig\ for a series of redshift steps that encapsulate the full range of the \zph-PDF as shown in the middle and bottom panels by the red solid dots.  Finally, we weigh the resulting SFR and \mstar\ distributions by the \zph-PDF probabilities and shown in the right-hand middle and bottom panels. For completeness, the horizontal dashed lines denote the resulting \cig\ fitted values for the {\it best-fit} \lephare\ input redshift value (i.e. a single redshift) with the corresponding \cig\ error estimates on these parameters shown as the vertical red error bars.
    }
    \label{fig:zpdf}    
\end{figure}

In this work we explore the SF-MS, focusing on using \cig\ \citep{noll09b}, which incorporates empirical relations to  build a grid of libraries that model: SFHs using SPS models \citep[e.g.][]{fiocroccavolmerange97,BC03,Maraston:2005MNRAS.362..799M}, interstellar gas and dust \citep[e.g.][]{calzetti94,calzetti97a,DH02}, fraction of AGN \citep[e.g.][]{fritz06} and optical and UV interstellar emission and absorption lines  \citep[see e.g.][]{kinney96,noll04}.  We use observations primarily derived from  the  VISTA Deep Extragalactic Observations (VIDEO) \citep{jarvis13}. 
\begin{figure}
\centering
 \includegraphics[width=0.49\textwidth]{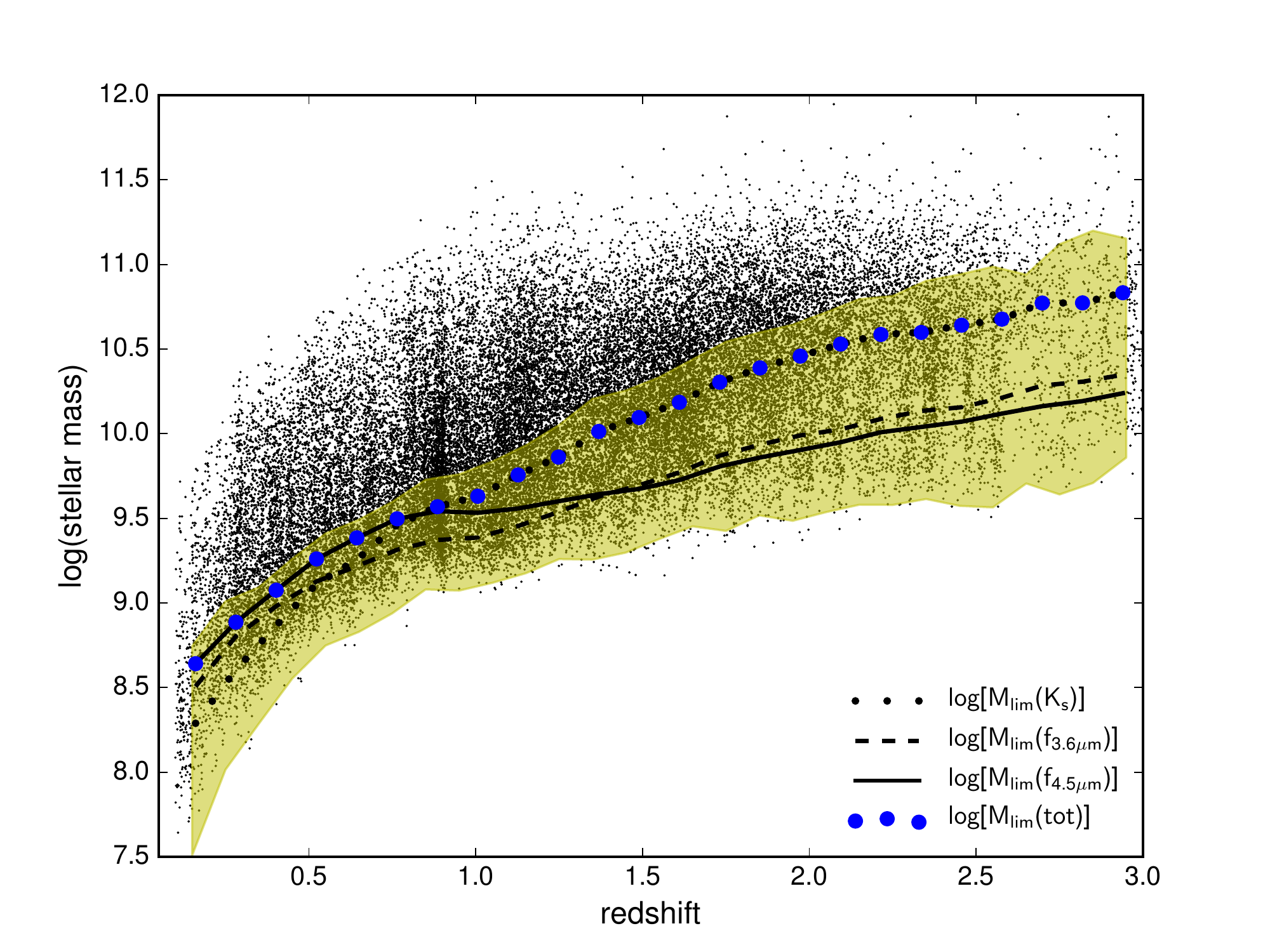}
     \caption{\small Stellar-mass completeness as a function of redshift.The black dots show the distribution our final sample of 52,812  galaxies. The transparent yellow region shows the stellar-mass limit $M_{\rm lim}^*$, for all objects that could be detected with an apparent magnitude limit of $K_{\rm s}<23.0$. The black dotted line indicates the mass completeness limit for the  $K_{\rm s}$ band. As our sample is jointly selected with IRAC1 and IRAC2 bands at 3.6 and 4.5\,$\mu$m we compute the mass completeness limits for these bands as shown by the dashed and solid black lines respectively.   The blue dots trace our final mass completeness limit which selects the highest mass limit from the three bands.}
    \label{fig:mcomp}    
\end{figure}

The format of this paper is as follows. In \S~\ref{sec:samp} we provide full details of the data and our sample selection followed by a description of the  SED fitting code \cig\  in \S~\ref{sec:cigale}. In this section we also discuss how we can extend this algorithm to account for the full propagation of photometric redshift PDFs and discuss completeness estimates and other sources of uncertainties.
In \S~\ref{sec:moddata} we discuss the various ways in which we model the SF-MS and then present our results in  \S~\ref{sec:results}. In \S~\ref{sec:sims} we compare our observational results to a variety of simulations to investigate how the key physical processes invoked in the simulations to reproduce local relations, effect the SF-MS at $0.5<z<3$.  A  final discussion and conclusions can be found in \S~\ref{sec:discussion}.  Throughout this work we adopt an H$_0$=70 km s$^{-1}$ Mpc$^{-1}$, $\Omega_M$=0.3, $\Omega_{\Lambda}$ = 0.7 cosmology.

\section{sample selection}\label{sec:samp}

\subsection{The VIDEO Survey}
For our analysis we use $K_s$-selected galaxies from the VISTA Deep Extragalactic Observations \citep[VIDEO; ][]{jarvis13} Survey. VIDEO is a $\sim12$~degree$^{2}$ survey in the near-infrared {\it Z, Y, J, H} and $K_{\rm s}$ bands that has been designed to trace the evolution of galaxies and clusters as a function of both epoch and environment  out to $z\sim4$.  VIDEO is being carried out over three of the most widely observed high-Galactic-latitude fields, covering $\sim$3~deg$^2$ in the ELAIS-S1 field, $\sim$4.5~deg$^2$ in the XMM-Newton large-scale structure field, and another $\sim$4.5~deg$^2$ in the extended Chandra Deep Field-South. These fields have a wealth of data from the X-rays through to the radio waveband, and are, along with COSMOS/UltraVISTA, the primary fields for observations with future facilities in the Southern hemisphere.
\begin{figure*}
\centering
 \includegraphics[width=0.48\textwidth]{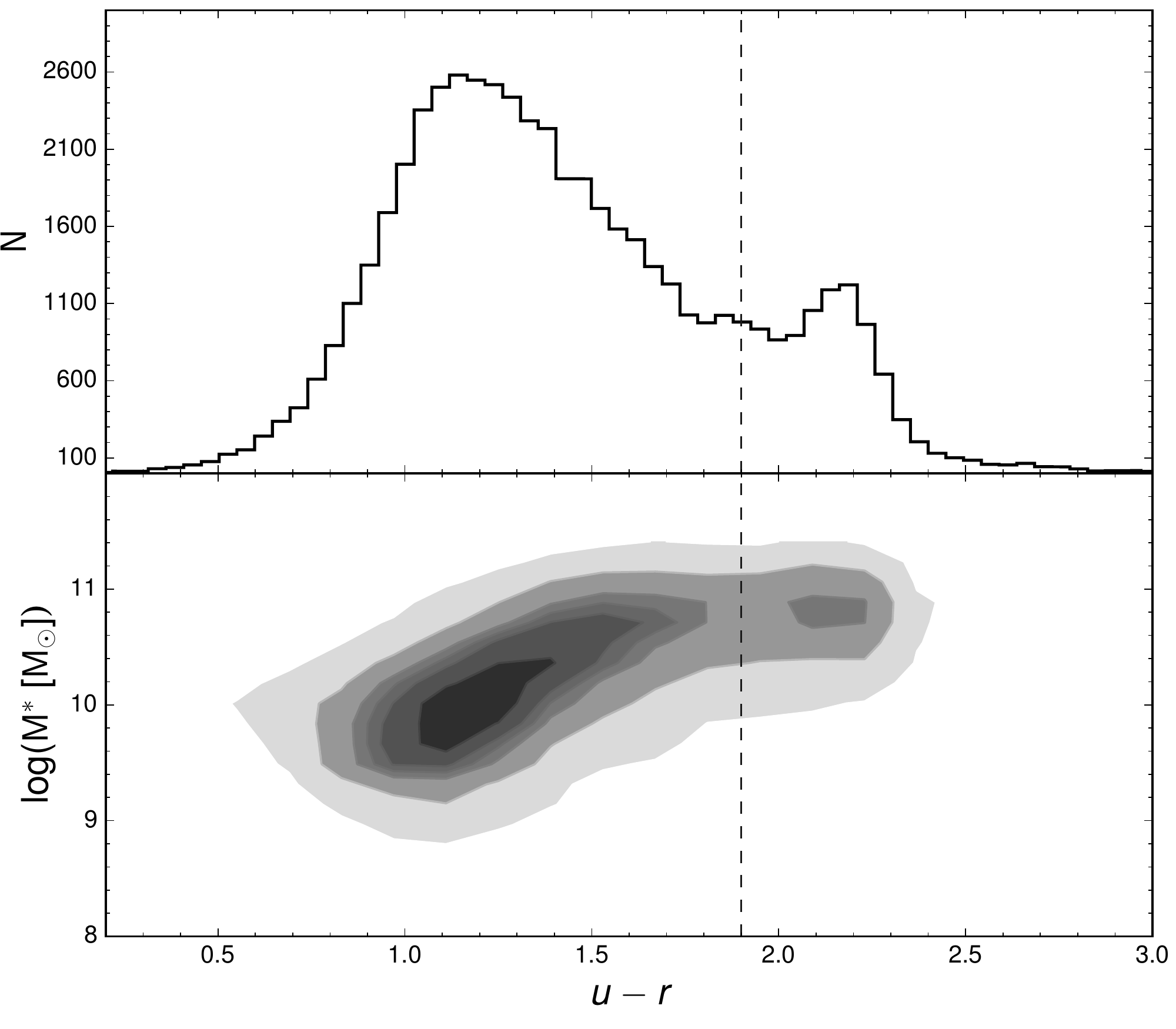}\hfill
 \includegraphics[width=0.48\textwidth]{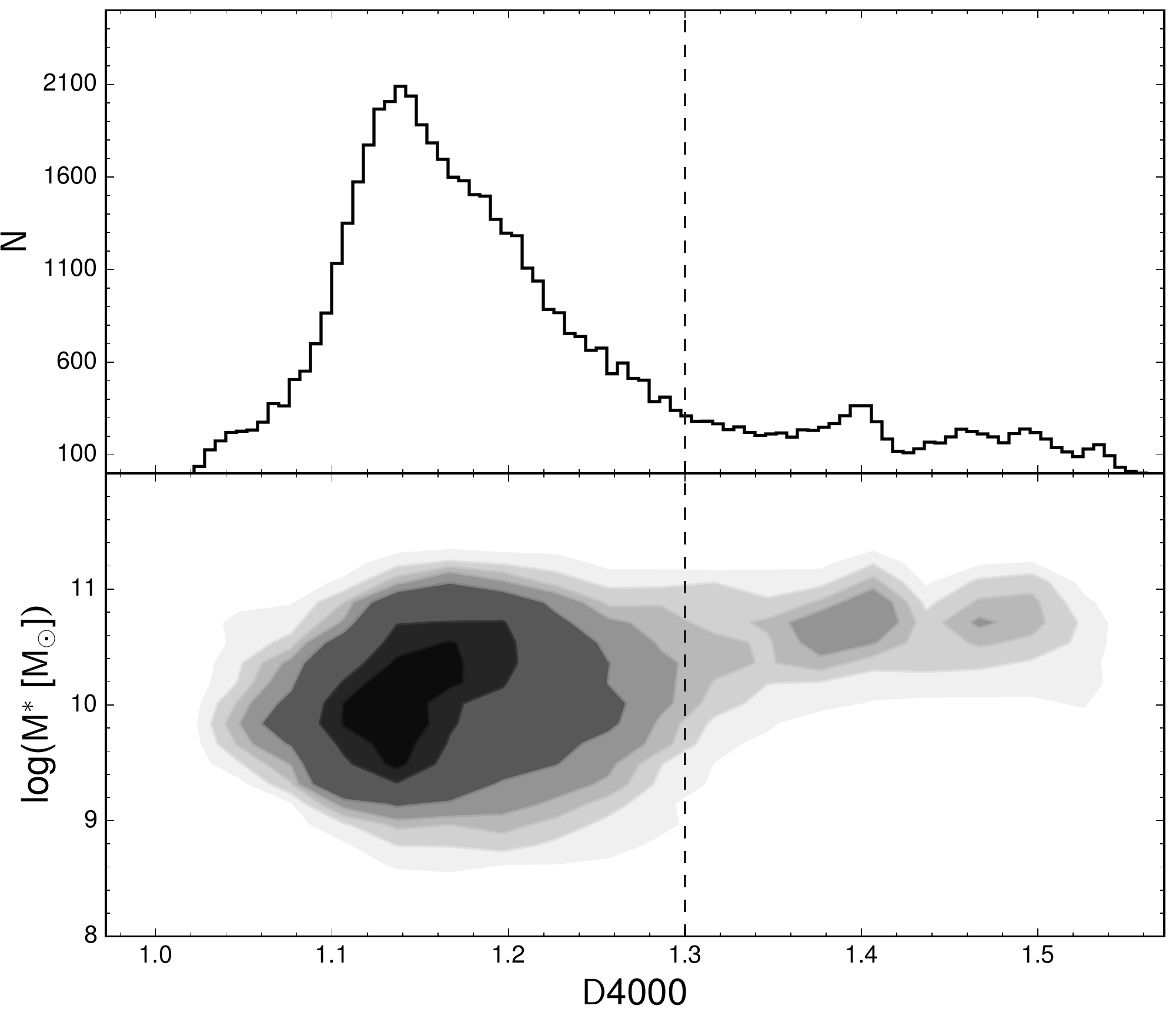}
 
     \caption{\small The two methods of selecting SFGs.  {\it Left panel}: the upper plot shows the rest-frame  \ur\ colour distribution. The bottom panel shows how this varies as a function of \lgmstar. The vertical dashed line indicates our colour cut to isolate SFGs; all galaxies with  \ur\ $>1.9$ are considered to be passive. {\it Right panel}: the D4000 index as determined by \cig. The extended tail in the D4000 distribution shows the residual passive galaxy population that we remove using a cut at D4000=1.3. The bottom panel shows the passive population are mostly massive objects with \lgmstar$\gtrsim 10$.}
    \label{fig:d4000hist}    
\end{figure*}

In this work we focus on an area within the XMM-LSS field where VIDEO observations overlap with the Canada-France-Hawaii-Telescope Legacy Survey Deep 1 \citep[CFHTLS-D1][]{Ilbert:2006AA...457..841I} field, providing optical photometry in the $ugriz$ bands. The creation of the CFHTLS/VIDEO catalogue extracted over the 10 bands ($ugriz$ from CFHTLS and $ZJHK_{\rm s}$ from VIDEO) is described in detail by \cite{jarvis13}. The current full catalogue contains 431, 949 objects over a $\sim 1$ deg$^2$ area.

To determine photometric-redshift (\zph) estimates with  VIDEO we used the public code \lephare\footnote{\url{http://www.cfht.hawaii.edu/~arnouts/LEPHARE/lephare.html}}\ \citep{Ilbert:2006AA...457..841I}. For more details regarding the \lephare\ settings see \citet{jarvis13}.  As is relevant to this work, \lephare\ outputs a redshift  probability distribution function (PDF) for each object, which we propagate in our analysis in deriving \sfr and \mstar\ estimates,
\begin{figure}
\centering
 \includegraphics[width=0.48\textwidth]{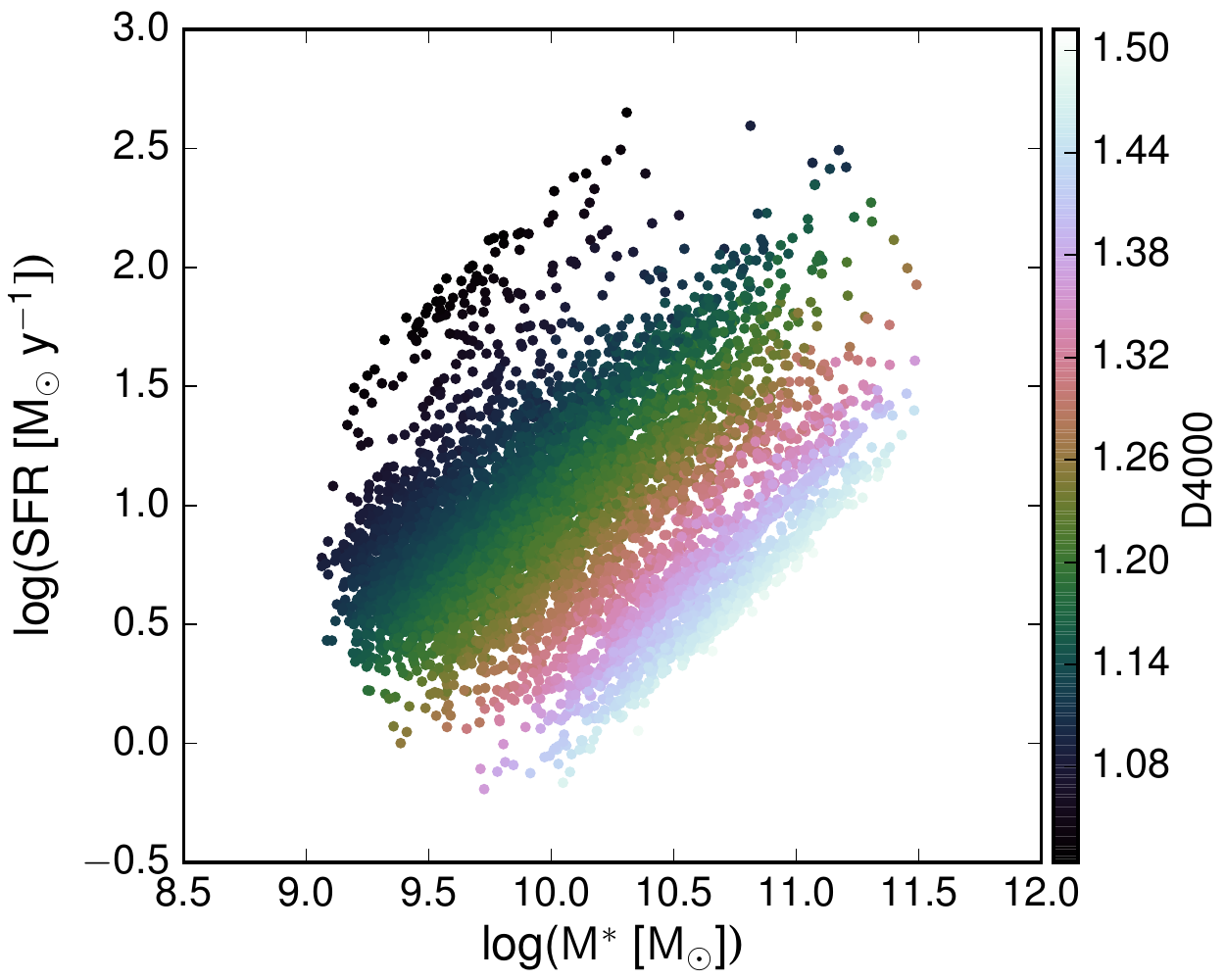}\hfill
     \caption{\small  Using the D4000 index to select star forming galaxies as discussed in \S~\ref{sec:sfselect}. The colour gradient shows how the D4000 index varies as function of  the \mstar-SFR distribution for a sample at $z\sim1$.  All parameters were derived from \cig.} 
    \label{fig:d4masssfr}    
\end{figure}

For our analysis we adopt a conservative approach and only select objects brighter than a limiting apparent $K_s$-band magnitude $K_{s}<23.0$, which corresponds to $>95$\,per cent completeness for the VIDEO data \citep[see ][]{jarvis13} for $0.1\leq z_{ph}\leq3.0$, thus reducing our parent sample to 74,063 galaxies.

\subsection{Matching to Multi-wavelength data}

The design of the VIDEO survey footprint was driven by the need to cover the wealth of ancillary data available in fields of several square degrees routinely accessible from Paranal. Multi-wavelength observations within VIDEO fields include {\em Spitzer} \citep[e.g.][]{Lonsdale:2003PASP..115..897L,mauduit12} and {\em Herschel} \citep{oliver12}, in addition to ongoing ground-based optical surveys from the CFHT, the VLT Survey Telescope (VST\footnote{\url{http://www.mattiavaccari.net/voice/ }}) and the Dark Energy Survey \citep{Flaugher:2005IJMPA..20.3121F,Bernstein2012}. For our analysis we thus built a well-sampled complete multi-wavelength dataset of galaxies detected by VIDEO and matched to available UV-to-IR  data. 

\subsubsection{SERVS Data}
The Spitzer Extragalactic Representative Volume Survey \citep[SERVS,][]{mauduit12} covers the three VIDEO fields and consists of deep observations in {\em Spitzer}-IRAC1 and IRAC2 bands at 3.6 and 4.5\,$\mu$m. We searched for SERVS counterparts to the VIDEO sources within a radius of 1" and only selected objects detected in both bands with fluxes higher than 1.0 $\mu$Jy (3-$\sigma$ limit). Our final sample contains 52,812 galaxies. This jointly selected sample at $K_{s}$, 3.6 and 4.5 \mic\  ensures that we can reliably detect the emission from the old stellar populations that contribute the bulk of the stellar mass, up to our redshift limit. 

\subsubsection{SWIRE and HerMES Data}

The {\em Spitzer} Wide-Area Extragalactic \citep[SWIRE][]{Lonsdale:2003PASP..115..897L} survey provides IRAC 5.8 and 8.0 \mic\  and MIPS 24, 70 and 160 \mic\ data over the whole of the VIDEO footprint. For SWIRE, we used the {\em Spitzer} Data Fusion (Vaccari et al. 2010\footnote{\url{http://www.mattiavaccari.net/df/}}) catalogues and matched SERVS and SWIRE positions within a search radius of 1".  

Far-infrared data was obtained  from the HERschel Multi-tiered Extragalactic Survey \citep[HerMES; ][]{oliver12} carried out with the SPIRE \citep{griffin10} instrument onboard the {\em Herschel Space Observatory} \citep{Pilbratt2010AA...518L...1P} at 250, 350 and 500 $\mu$m. The SPIRE source extraction is based on forced photometry at the IRAC positions of 24 $\mu$m sources as described in \citet{roseboom10}, and using an input catalogue based on the {\em Spitzer} Data Fusion. As with the SWIRE data we match the \citet{roseboom10} catalogue to our VIDEO-SERVS  objects within 1\,arcsec.

\section{Estimating {\it SFR} and {\it Mass} with \cigbig}\label{sec:cigale}

To estimate SFRs and \mstar\ we use the public SED fitting code, {\it C}ode {\it I}nvestigating {\it GAL}axy {\it E}mission, \citep[\cig\footnote{\url{http://cigale.lam.fr/}}; ][]{burgarella05b,noll09b,giovannoli11}, which is optimised to provide physical information of galaxies by fitting their UV-to-IR SEDs.  In this section we summarise the main steps of the algorithm.

\cig\ is based on the use of a UV-optical stellar SED plus an IR-emitting, dust component. In essence, \cig\ builds  UV-to-IR models which are then fitted to the observed SEDs. The code  fits the observed data in the UV, optical, and NIR with models generated with a stellar populations synthesis code, assuming a star formation history (SFH) and a dust attenuation as input.  The energetic balance between dust-enshrouded stellar emission and re-emission in the IR is  conserved by combining the UV/optical and IR SEDs.

In our implementation, we have used the  stellar population synthesis models of  \cite{Maraston:2005MNRAS.362..799M} (hereafter M05)  which consider the thermally pulsating asymptotic giant branch (TP-AGB) stars, and use the \cite{kroupa01} initial mass function (IMF). We generate star formation histories (SFH) based on two stellar populations: a recent stellar population with a constant SFR on top of an older stellar population created with an exponentially declining SFR. 
Model spectra are then reddened using attenuation curves from \cite{calzetti00}, and we use the  semi-empirical one-parameter models of \citet{DH02} (hereafter DH02) to fit the stellar-heated dust emission.  The DH02 library is composed of 64 templates parametrised by $\alpha$, the power-law slope of the dust mass over heating intensity. $\alpha$ is directly related to the $f_{60\mu m}$/$f_{100 \mu m}$ flux ratio, where $f_{60\mu m}$ and $f_{100\mu m}$ represent fluxes at 60 and 100\,$\mu$m, respectively. \cig\ allows us to estimate the fraction of  ${L_{\rm IR}}$ due to an AGN by combining AGN with the starburst templates.  We used the set of six templates from the \citet{fritz06} library. The final part of this process uses the \cite{kinney96} and \cite{noll04} spectra  to create empirical templates  to consider the UV and optical interstellar emission and absorption lines respectively. The SFRs are computed as 
\begin{equation}
{\rm SFR}=\frac{M_{\rm gal}}{\tau(e^{t/\tau}-1)},
\end{equation}
where $M_{\rm gal}$ is the galaxy mass, $\tau$ is the $e$-folding (or decay) time and $t$ is the look-back time.
Finally, the physical parameter values corresponding to the best-fit model are constrained via a  $\chi^2$ minimisation.

We explore the impact of the \cig\ outputs when flux information is unavailable in certain filters for a given galaxy. The fraction of objects in our sample that were detected in IRAC 3 \& 4, MIPS and SPIRE is quite small. Thus, the dominating contributing filters used in \cig\ are from VIDEO, CFHT and SERVS.  For this test  we have isolated those galaxies that have detections across the entire filter set. We then run \cig\ twice;  first with input photometry from  {\it only} VIDEO, CFHT and SERVS, and secondly where we add the additional information from IRAC 3 \& 4, MIPS and SPIRE.  Figure~\ref{fig:mips} shows the  comparison between the estimated SFR and \mstar\  from \cig\ under these two scenarios.
For the comparison of SFRs, we find very good agreement between the outputs, with an average overall scatter in the residuals of $\sigma_M=0.22$ dex.   
Similarly for \mstar\, we find a very tight correlation with an average scatter of just $\sigma_M=0.07$ dex in the residuals.  We therefore conclude that \cig\ can robustly estimate the SFR and \mstar\ without the need for the far-IR data to the redshift limit of our sample.
\begin{table}
\caption{ Input parameter for propagating the \zph\ PDFs through \cig. The first column shows the range of 
ages for the stellar population (SP) models used to create star formation histories. The second column shows the redshift range we permit based on the SP ages.   }
\label{tab:cig}
\centering
\begin{tabular}{c c c}
Ages of old SP models & Data z-range  & z-step coverage \\
$\tau_f$ (Gyr) & (best-fit \zph) & (\cig\ model range)\\
\hline\hline
5, 6, 7, 8, 9                   & $0.1 <  z \leq 0.3$   & $0.01 < z_{step}  < 0.41$ \\
2.5, 3.5, 4.5, 5.5, 6.5  & $0.3 <  z \leq 0.6$   & $0.15 < z_{step}  < 0.75$\\
1, 2, 3, 4, 5                   & $0.6 <  z \leq 1.0$  & $0.45 < z_{step}  < 1.15$\\
1, 2, 3, 4                       & $1.0 <  z \leq 1.4$  & $0.85 < z_{step}  < 1.55$\\
1, 1.5, 2, 2.5, 3            & $1.4  <  z \leq 2.0$ & $1.25 < z_{step}  < 2.15$\\
1, 1.5, 2, 2.5                 & $2.0<  z \leq 2.4$   & $1.85 < z_{step}  < 2.55$\\
1.0, 1.2, 1.5, 1.7,1.9    & $2.4  <  z \leq 3.0$ & $2.25 < z_{step}  < 3.15$\\
\hline
\end{tabular}
\end{table}

\subsection{Propagating the \zph\ uncertainties through \cig}\label{sec:zpdfcigale}
\cig\ requires the redshift of the galaxy as input and  is not currently optimised to simultaneously constrain redshift {\em and} physical properties of the galaxies. We therefore modified  \cig\ to obtain both SFR and \mstar\ estimates when we have the full photometric  redshift PDF.  For our sample such information was obtained from \lephare\,  and propagating this through the \cig\ framework required only a slight modification.

This modification is illustrated in Figure~\ref{fig:zpdf}. The top panel shows a typical redshift PDF outputted  from \lephare, where the blue circles represent the resolution of the \lephare\ grid for a given galaxy.  As we can see in this example, there are two peaks in the \zph-PDF; the first at $z$$\sim$0.3 and the second at $z$$\sim$0.6 with a lower probability. The majority of the contribution is centred  at $z$$\sim$0.3  and \lephare\   returns a best-fit value as indicated by the vertical dashed line. To include the full \zph\ contribution for a given galaxy we run \cig\ at a series of redshift steps that cover the range that fully describes the \zph-PDF. The middle and bottom left-hand panels in Figure~\ref{fig:zpdf} show the resulting $\log_{10}$(SFR) and $\log_{10}(M_*)$ estimates from \cig\ respectively as a function of redshift. Finally, we can weight this relation  in SFR and \mstar\ according to the interpolated probability distribution of the \zph, as shown in the right-hand  bottom and middle panels. We also show the estimated error on the SFR and \mstar\ from \cig\ for this galaxy using the best-fit redshift value.

To account for the range in the full photometric redshift PDF estimation from \lephare, we ran \cig\ with different stellar population (SP) model ranges according to a set of redshift slices which are summarised in Table~\ref{tab:cig}.  The use of applying two stellar populations was explored in more detail recently by  \cite{Buat:2014AA...561A..39B}, who applied a similar approach to derive SFRs and \mstar\ estimates at redshifts $z>1$. Our approach  allows for \zph-PDFs that may not be well described by e.g a simple gaussian. In \S~\ref{sec:results} we apply this approach in determining SF-MS.

\subsection{Stellar mass completeness}
To estimate the  completeness limit for \mstar\ we follow the approach used by \cite{Ilbert:2013AA...556A..55I} and \cite{Pozzetti2010AA...523A..13P}, and which is illustrated in Figure~\ref{fig:mcomp}, where we show the stellar masses for our sample of 52,812 galaxies as a function of redshift. We calculate the lowest stellar mass which could be detected for a given galaxy with 
\begin{equation}
\log_{10}(M_{\rm lim}) = \log_{10}(M_*)+0.4(K_{\rm s}-K_{\rm s}^{\rm lim}),
\end{equation}
where, in our study, the apparent magnitude limit, $K_{\rm s}^{\rm lim}= 23.0$.  This is shown as the transparent yellow region on Figure~\ref{fig:mcomp}. We then bin the data in redshift and compute the 90th percentile of stellar mass within each bin, shown as blue dots in the figure. Thus for a given redshift, we can interpolate the  stellar mass completeness limit that corresponds to the mass for which 90 per cent of galaxies fall below the percentile. 
\begin{figure}
\centering
 \includegraphics[width=0.5\textwidth]{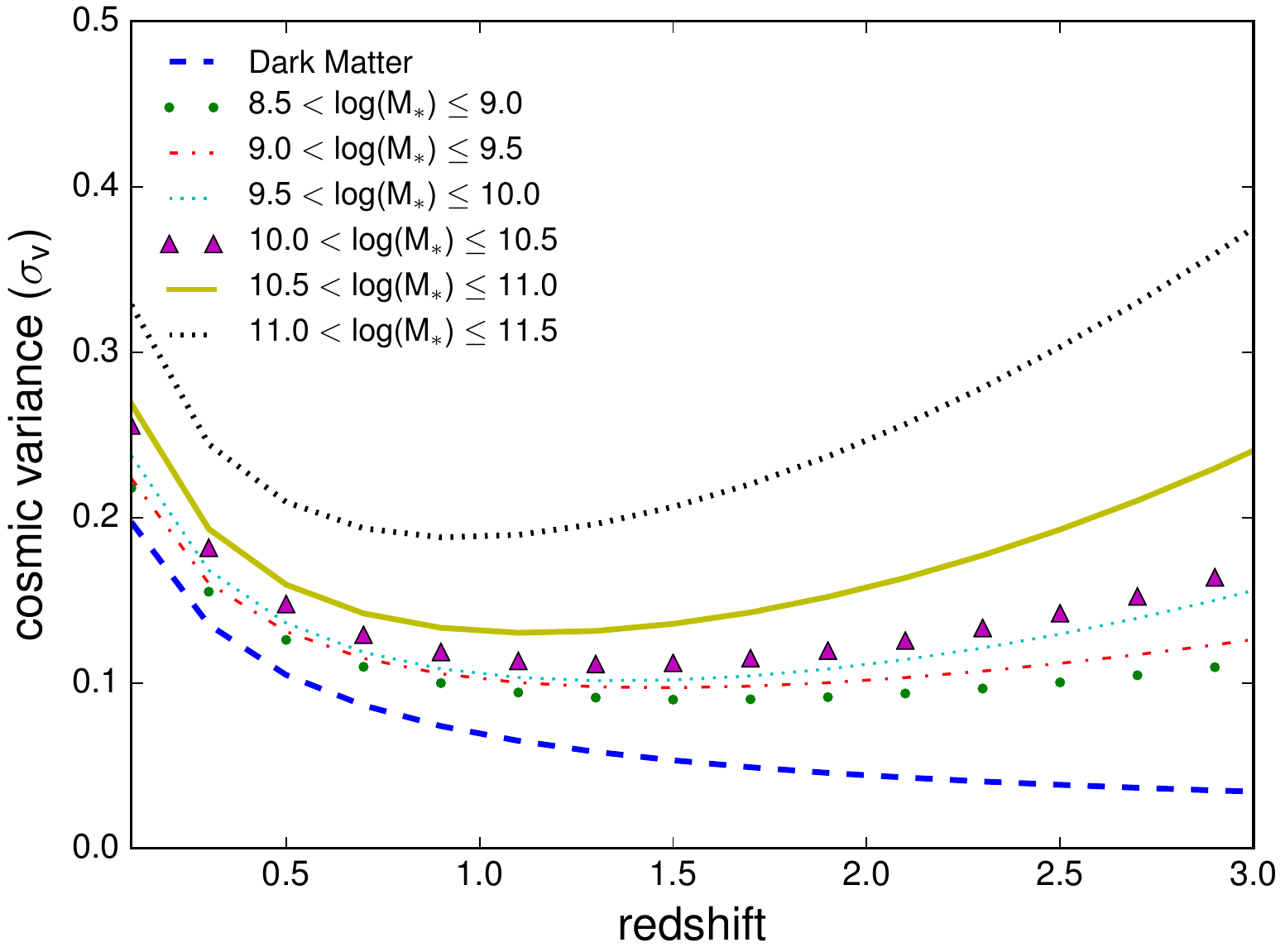}
     \caption{\small Cosmic variance as a function of redshift for a range of mass bins for the current VIDEO data release covering 1~deg$^2$.   }
    \label{fig:sigcv}    
\end{figure}

\begin{figure*}
\centering
 \includegraphics[width=1.02\textwidth]{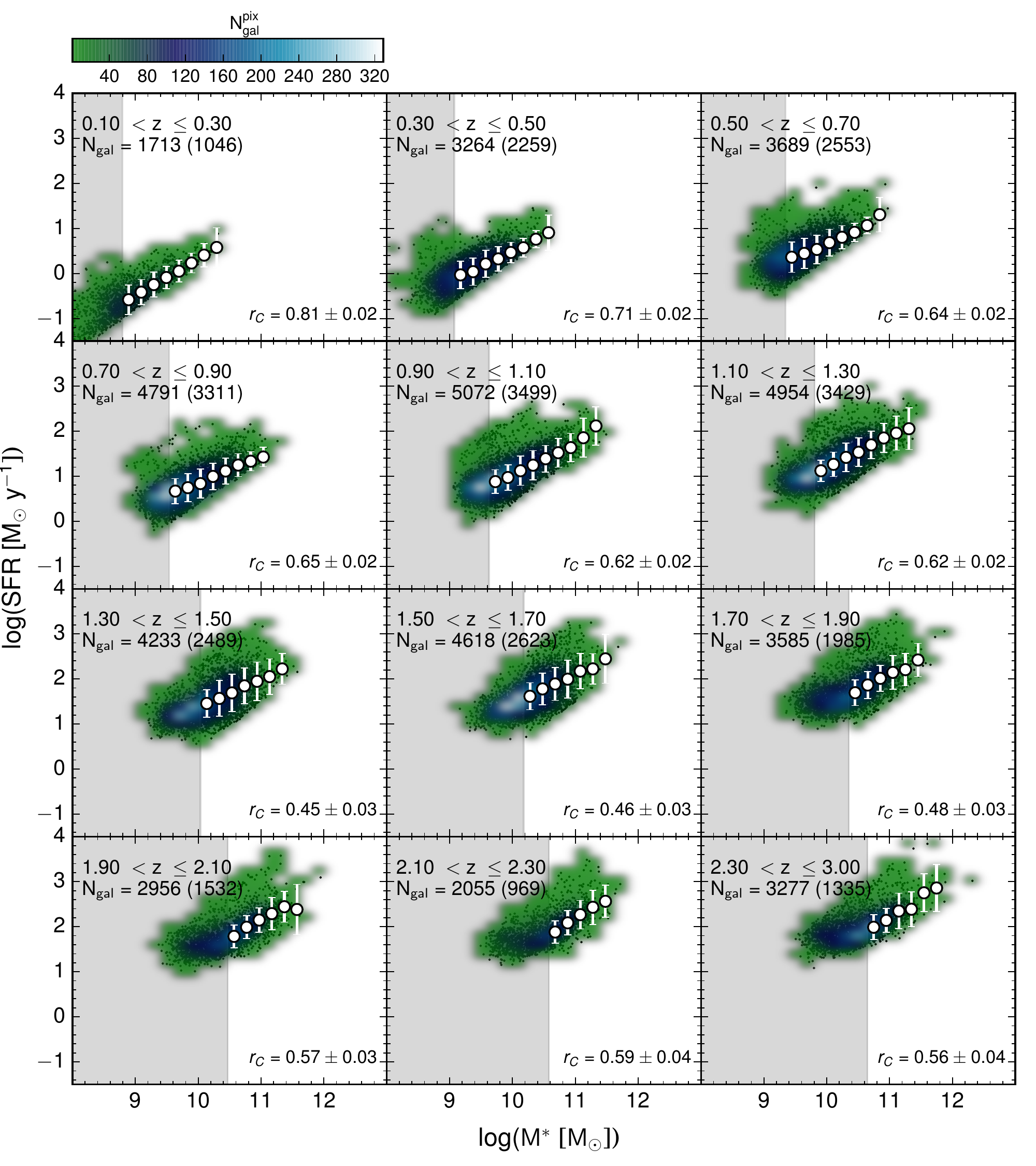}
     \caption{\small The SF-MS for SFGs from  $0.1<z<3.0$. The coloured regions are a 2D histogram of the  SFR-\mstar\ distribution, where the number of objects in each pixel ($\delta{\rm SFR}=\delta$\mstar= 0.2) is indicated in the colour bar at the top of the figure ($N^{\rm pix}_{\rm gal}$). The white filled circles denote the median values with their respective uncertainties computed as  \errtot=$\sqrt{(\sigma_{\rm MAD}^2+\sigma_{\rm V}^2)}$ which combines the mean absolute deviation and cosmic variance (see text for details). The grey area shows the region of our stellar mass completeness limit. Any data within this region is not considered in the final analysis. $N_{\rm gal}$ indicates the total number of galaxies in each panel, with the total number within our stellar mass completeness range shown in parenthesis.  The correlation coefficient between SFR and $M_*$  for the stellar mass-complete data is shown as $r_c$ in each panel.  }
    \label{fig:vidmedzbest}    
\end{figure*}

\begin{figure}
\centering
 \includegraphics[width=0.49\textwidth]{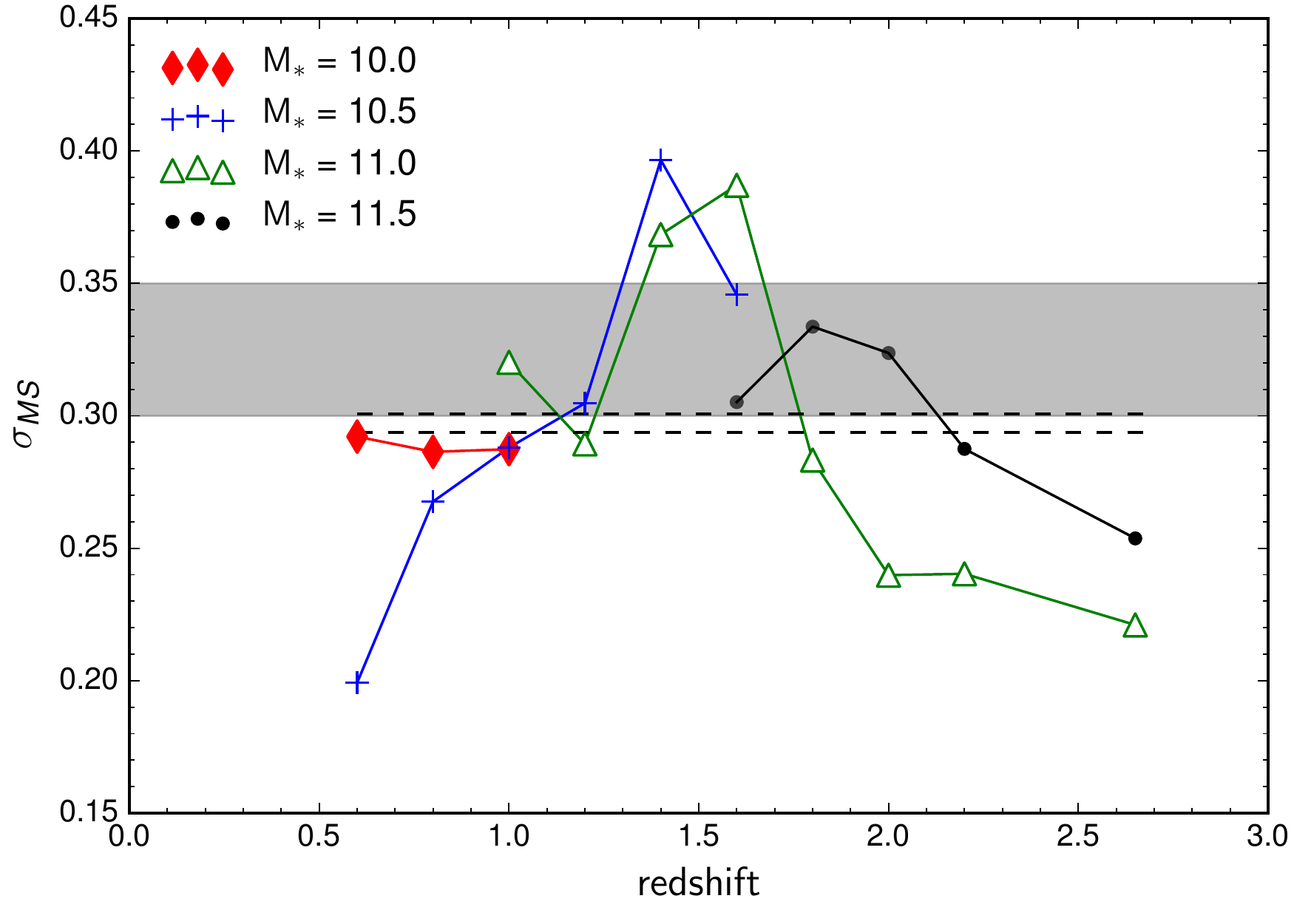}
     \caption{\small  The dispersion in the SF-MS,  \sigms,  as a function of redshift for a series of stellar-mass bins. The grey shaded region shows a range of \sigms\ as reported  in e.g. \citet{Noeske:2007ApJ...660L..47N,Whitaker2012} and \citet{Magnelli2014AA...561A..86M}.  The region between the dashed line indicates our average dispersions across all stellar mass bins where for \lgmstar=10.0, 10.5, 11.0 and 11.5,  \sigmsmean = 0.29, 0.29, 0.31 and 0.28 respectively as estimated within the stellar mass completeness limits. }
    \label{fig:sigmams}    
\end{figure} 
\subsection{Selecting star-forming galaxies}\label{sec:sfselect}
To analyse the SF-MS we need to isolate SFGs and remove any passive galaxies that would contaminate the sample.  However, selecting SFGs from multi-wavelength data can be done in a variety of ways. This was recently reviewed by \cite{Speagle2014ApJS..214...15S} and explored in detail by \cite{Rodighiero2014MNRAS.443...19R}, where they highlighted how different colour selections can lead to different SFG populations being isolated, contributing to  the broad range of derived slopes in the literature. For example, applying a {\it BzK}  selection \citep[e.g.][]{daddi04,Rodighiero2011ApJ...739L..40R,Kashino2013ApJ...777L...8K} tends to isolate more active SFGs, biasing against galaxies which would  otherwise  be classified as SF by other methods (e.g. NUV$rJ$). This type of selection generally leads to steeper derived slopes for the SF-MS. To help avoid selecting only the bluest most active galaxies, \cite{Ilbert:2013AA...556A..55I} used a mixed approach and applied two colour selections in $r-J$ and NUV-$r$, which provided a broader range of SFGs.  In a recent paper by \citet{Renzini2015ApJ...801L..29R} they suggest a construction of a 3D SFR-Mass-Number distribution may help in defining a more objective SF-MS by adopting a cut at the ridge line of the star-forming peak.

In this study we explored two approaches which yielded very different results in our final analysis. With the filter coverage available to us we found a rest-frame  \ur\ was the optimal colour selection approach, but not necessarily the most effective way to remove the passive population. The top-left panel of Figure~\ref{fig:d4000hist} shows the \ur\ distribution where we observe a bimodal distribution, split approximately at \ur\ $\sim1.9,$ as indicated by the vertical dashed line. The bottom-left panel shows the   \ur\ as function of stellar mass (as determined using \cig) which perhaps better illustrates this segregation in the two  populations. As such we removed all objects in our final sample with \ur\ $>1.9$. However, as we shall discuss in the \S~\ref{sec:results}, this does lead to a significant fraction of passive galaxies being missed, along with potential SFGs being removed from the sample at redshifts $z\lesssim 1$.

Alternatively, one may use the 4000\,\AA\ break (hereafter referred to as D4000).  The D4000 index provides useful information regarding the age of the stellar population and perhaps provides a more physical selection process. A low index indicates a young stellar population, whereas the larger the value the older the galaxies \citep[e.g.][]{kauffmann03a,Westra2010ApJ...708..534W}. We use \cig\ to determine the D4000 index as part of the fitting procedure, adopting the  definition as detailed in \cite{Balogh1999ApJ...527...54B}. 
Moreover, the calculation is intrinsically a dust-free measurement, as it is based on the unreddened spectra. Whilst this approach is SED-model dependant, it does provide us with a self-consistent approach to identify SFGs.
In the right-hand panel of Figure~\ref{fig:d4000hist} we show the resulting D4000 distribution  as a function of \mstar. We apply a conservative cut at  D4000 = 1.3, and consider galaxies beyond this limit to be older passive galaxies. For the remainder of this article our results will be based on this D4000 selection. As we can see in Figure~\ref{fig:d4masssfr}  this type of selection imposes a much sharper cut in the SFR-\mstar\ plane as a function of redshift impacting on conclusions drawn about the evolution of the SF-MS as we will discuss in more detail  in \S~\ref{sec:sfsel}.

\section{Modelling the SF-MS}\label{sec:moddata}
 We parameterise and model the SF-MS in a number of ways to test the robustness of our findings and to allow direct comparison with a number of related works. The most common approach is to assume a simple power-law of the form
\begin{equation}
\log_{10}({\rm SFR}) =  \alpha\log_{10} (M_*)+\beta \label{equ:powerlaw},
\end{equation} 

\citep[see e.g.][]{Noeske2007ApJ...660L..43N,daddi07,Elbaz2007AA...468...33E,Santini2009AA...504..751S,Heinis:2014MNRAS.437.1268H}.  Without a redshift dependent term, the data can be simply split into a series of redshift slices, and a fit is perfomed within each slice to determine how the normalisation $\beta$ and the slope $\alpha$ of the SF-MS evolve.

We calculate the total error on the median SFRs in each bin as \errtot=$\sqrt{(\sigma_{\rm MAD}^2+\sigma_{\rm V}^2)}$, where $\sigma_{\rm MAD}$ is the error on the median (median absolute deviation, or MAD), and is related to the standard error on the mean $\sigma$ by $\sigma<x>$ = $\sigma\bar x/1.4826$ \citep{Hoaglin:1983ured.book.....H}. This is  combined in quadrature with the  cosmic variance uncertainty $\sigma_{\rm V}$. 
This is the uncertainty in observed number density of galaxies arising from the underlying large-scale density fluctuations.  We have used the public code \getcv\ developed by  \cite{moster:2011ApJ...731..113M}, to estimate cosmic variance as a function of redshift for the range in \mstar\ considered.   Figure~\ref{fig:sigcv} shows how the fractional cosmic variance error \sigv\ varies as a function of redshift for a range of stellar  mass bins as computed for our  sample.  We then perform a least-squares fit to estimate the power-law parameters  in Equation~\ref{equ:powerlaw}.  

Although using  the median SFR can be useful when dealing with a statistically-small sample, it naturally leads to a loss of information.  Therefore, we  also perform power-law fits within each redshift slice to the {\it full} galaxy distribution which we refer to as `all-data' fits and consider the uncertainty on the SFR and \mstar\ for each individual galaxy as determined using \cig.    

 \begin{figure*}
\centering
 \includegraphics[width=0.48\textwidth]{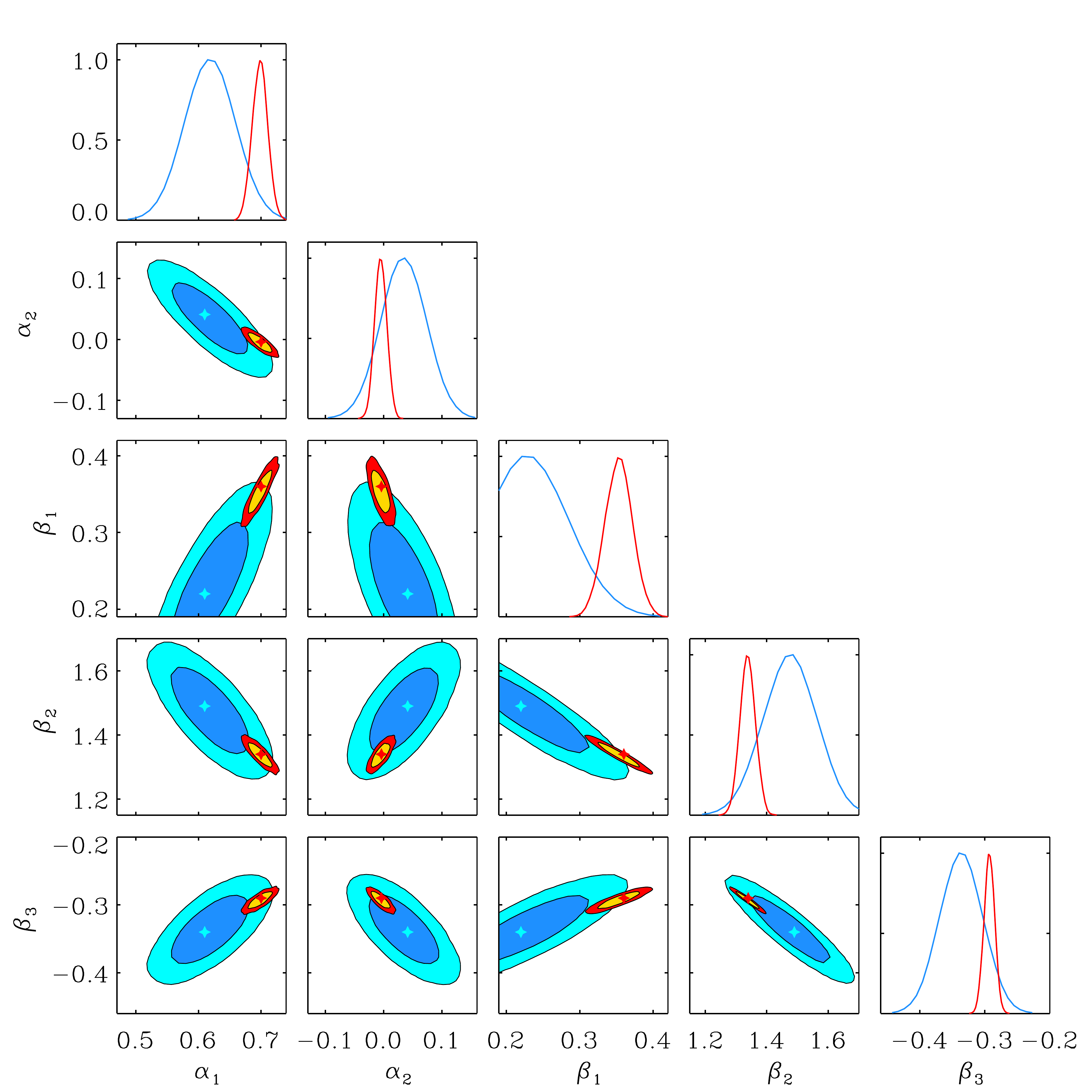}\hfill
 \includegraphics[width=0.48\textwidth]{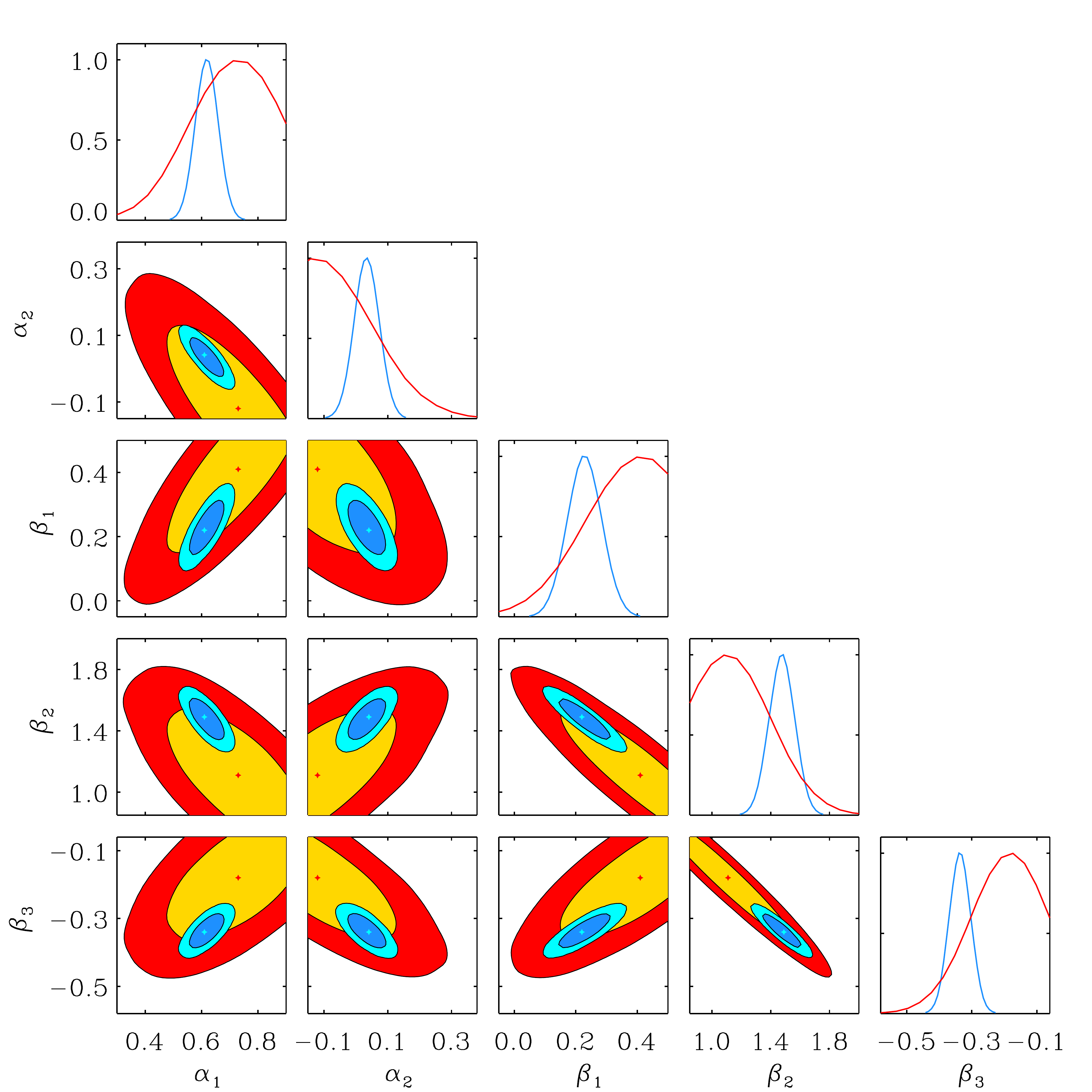}
     \caption{\small  Parameter constraints from our MCMC analysis of the model shown in Equation~\ref{equ:W12}.  {\it left panel set:}  The dark and light blue shaded contours show the 2-D 1$\sigma$ and 2$\sigma$ marginalised results respectively. The red and yellow contour regions show an example from the 1000 Monte Carlo  \zph-PDF  constraints. The 1-D marginalised probability distributions are shown on the diagonal as the blue line (red line) for the VIDEO median (\zph-PDF) fits.  The  cross near the centre of each contour set shows the best-fit parameter value.  {\it right panel set:}  Comparison of   \citet{Whitaker2012} median fits (red and yellow shaded contours) with our median fits (blue contours). }
    \label{fig:mcmc}    
\end{figure*} 
Recent studies by e.g.  \cite{Magnelli2014AA...561A..86M,Whitaker2014ApJ...795..104W,Schreiber2015AA...575A..74S} and \cite{Tasca2014arXiv1411.5687T}  have  found a flattening of the MS towards  the high mass end, and therefore require more robust modelling using either e.g. a broken power-law or   an extra term in the form of a quadratic given by
\begin{equation}\label{equ:magnelli}
\log_{10}(\rm SFR) = A_1 +A_2\log_{10}(M_*)+A_3[\log_{10}(M_*)]^2.
\end{equation}
Others have instead chosen to model the SF-MS jointly with a redshift-evolution term. 
In \cite{Schreiber2015AA...575A..74S} they parameterise as
\begin{align}\label{equ:schreiber}
\log_{10}[{\rm SFR(z)}]&=\log_{10}(M_*)-\alpha_1 +\beta_1\log_{10}(1+z)  \nonumber \\
	&-\beta_2[\max(0,\log_{10}(M_*)-\alpha_3\log_{10}(1+z)]^2
\end{align}
Alternatively,  \cite{Whitaker2012}, applied  a second order polynomial of the form
\begin{align}
\log_{10}[{\rm SFR(z)}] &=\alpha(z)[\log_{10}(M_*) -10.5] +\beta(z), \label{equ:W12}
\end{align}
where  the terms $\alpha(z)$,  $\beta(z)$ are given by,
\begin{align} \nonumber
\alpha(z) &= \alpha_1 + \alpha_2z, \ \mbox{and} \\\nonumber 
\beta(z)&=  \beta_1 + \beta_2z +  \beta_3z^2,
\end{align}
and  $\alpha_1$, $\alpha_2$, $\beta_1$, $\beta_2$ and $\beta_3$ are the fitting parameters.  To explore evolution we adopt the \cite{Whitaker2012} form, applying our \zph-PDF approach as detailed in \S~\ref{sec:zpdfcigale}. In practice  we create 1000 Monte Carlo (MC) samples randomly drawn from these PDFs, which are then individually fitted via Monte Carlo Markov Chain (MCMC). We can then examine the distribution of parameter constraints to assess any bias from the standard fitting procedure. We have used the public  MCMC algorithm  \cosmo\ \citep{cosmomc}.  Samples are generated using the Metropolis--Hastings algorithm  \citep{hastings:1970} for each of several intercommunicating Markov chains, with a `proposal distribution function' that allows full exploration of the posterior space.  We adopt a $\chi^2$ likelihood given by
\begin{equation}\label{equ:like}
-2\log_{10}(\mathcal L) \equiv \chi^2=\sum_{i=1}^N\frac{[{\rm \log_{10}(SFR^{data}_i)-\log_{10}(SFR^{mod}_i)}]^2}{\sigma_{\rm tot, i}^2}.\end{equation}
where $\log_{10}({\rm SFR}^{{\rm data}}_i)$ is the observed data and $\log_{10}({\rm SFR}^{\rm mod}_i)$ is the model. We apply this model to our \zph-PDF analysis of the SF-MS and our  \errtot\ in this case includes the intrinsic dispersion of the SFR-$z$ and \mstar-$z$ distributions  as well as \sigv\, giving $\sigma_{\rm tot}(z)$=$\sqrt{[\sigma_{M_*}(z)^2+ \sigma_{\rm SFR}(z)^2+\sigma_V^2]}$. 

\subsection{Calibration of the main-sequence}\label{sec:cal}
In \S~\ref{sec:plfits} we compare our SF-MS results to  a range of recent studies.  However, as we can see from Table~\ref{tab:assumps}  there is a spread in the underlying assumptions to determining   stellar masses and SFRs that include: wavelength coverage/selection of an observed sample, SFR estimation (\lsi\ relation), the  IMF,  stellar population synthesis models (SPS), extinction, metallicities,  adopted cosmology, not to mention other contributing factors such as the modelling star formation histories (SFH),  dust attenuation, photometric redshifts and incompleteness. Whilst for any given study the results may be self consistent, quantifying differences in the SF-MS  beyond their intrinsic scatter due to differing  assumptions can be challenging. In general, calibration of MS data is typically an average correction which impacts the normalisation. In an attempt to  make a fair comparison we have adopted a similar  calibration approach detailed in \cite{Speagle2014ApJS..214...15S} (S14).  In their study they examined twenty-five recent studies and highlight  three major sources that require a  calibration offset,  namely  the choice of IMF, SFR indicators (\lsi\ relation) and  SPS modelling.  For the works that overlap  with ours in Table~\ref{tab:assumps} we have used the S14  re-calibrated MS power-law fits which we show in Table~\ref{tab:fig11fits}.  
Whilst S14 estimate offsets for e.g. adopted cosmology, extinction, emission line, these were shown to be sub-dominant to the IMF, SPS models and SFR indicators. As such, in  our calibration we consider only contributions from the latter. 

The IMF offsets used in S14 were taken from taken from \cite{Zahid2012ApJ...757...54Z} where they calibrated to the  Kroupa IMF using
\begin{equation}
M_*^K =1.06M_*^C =0.62M_*^S,
\end{equation}
where the superscripts refer to Kroupa, Chabrier \citep{chabrier03}, and Salpeter \citep{salpeter55} IMFs, respectively. As they calibrate to the Kroupa IMF, these would lead to respective  offsets in stellar mass for Chabrier and Salpeter  of  $C_{M_*}^C$ = +0.03 and $C_{M_*}^S $= -0.21.  This would translate to an offset in the normalisation of the MS fit  by 
\begin{equation}\label{equ:betacal}
\beta_{C}=\beta_0-\alpha\times C_{M_*},
\end{equation}
where $\beta_C$ is the calibrated normalisation, $\beta_0$ is the original fitted normalisation and $\alpha$ is the slope. Since we have applied  the Kroupa IMF in our analysis we do not include this offset in our calibration.

 There have been a number of investigations  into the differing approaches to generating SPS models and in particular the  treatment of the modelling of thermally pulsating asymptotic giant branch  (TP-AGB) stars.   \citep[e.g.][]{Salimbeni2009AIPC.1111..207S,Conroy2009ApJ...699..486C,Magdis2010MNRAS.401.1521M,Walcher2011ApSS.331....1W}. To calibrate from M05 to BC03  (as in S14) we adopt the conservative offset of  $C_{M_*}^{SPS}$=~+0.15~dex which was an estimated average from analyses from \cite{Magdis2010MNRAS.401.1521M,Salimbeni2009AIPC.1111..207S,Conroy2009ApJ...699..486C}. Therefore, we apply  Equation~\ref{equ:betacal} to obtain the MS  offset. 

In a recent study by \cite{Buat:2014AA...561A..39B} they examined potential SFR offsets derived from  \cig\ compared to that of a standard \lsi\ relation that convert FUV and IR luminosities in to SFRs,
and found a very tight correlation with that of the \cig\  derived SFRs.  In the case of the two population model, which we have applied in our analysis, they reported a small offset  resulting in a slight underestimate of the instantaneous SFR with  ${\rm SFR_{IR,FUV}}$. For our analysis we therefore do not attempt a correction, but intend to investigate this more deeply in subsequent work.

\begin{table*}
\caption{Details of the properties of the various constraints we compared to in Figure~\ref{fig:sfrmasscomp}. Column 2: the initial mass function (IMF) used, column 3 is the  indicator used estimate SFRs, column 4:
wavelength selection of the parent sample, column 5: method to select star-forming galaxies, column 6: stellar population synthesis (SPS) model used, column 7: extinction law, column 8: adopted cosmology, column 9: survey and column 10: total area
of sample. }
\label{tab:assumps}
\centering
\scriptsize
\begin{tabular}{c c c c c c c c c c}
Source      & IMF 	 &  SFR Indicator 	& Wavelength   & SFG  		& SP Model & Extinction & Cosmology & Survey                            & Area  \\
		&		&				& selection	& selection 	&		    &	Law       &($h,\Omega_m,\Omega_\Lambda)$ & & deg$^2$\\
\hline\hline

This work         							& Kroupa 		& SED (\cig)     		& $K$/3.6, 4.5\,$\mu$m		& D4000, {\it u-r}   & M05      	&  $^d$C00		& (0.7,0.3,0.7) 		& VIDEO/CFHT/SERVS   		 & 1       	\\
\citet{daddi07}  						      	& Salpeter 	& H$\alpha$      	& $K$				        & {\it sBzK}	    & BC03        	& $^e$P84		& (0.7,0.3,0.7) 		& GOODS                         		&  0.033 	\\ 
\citet{Dunne:2009MNRAS.394....3D}     		& Salpeter 	& 1.4 GHz  		& $K$					& None		    & BC03 	&  C00			& (0.71,0.27,0.73)	& UDS/UKIDSS 			&  0.8 	\\
\citet{Elbaz2007AA...468...33E} 			& Salpeter 	& H$\alpha$ 		& UV/opt					& blue		    & $^a$PEGASE.2	&  $^f$CF00 		& (0.7,0.3,0.7)		& SDSS/GOODS			&  705 	\\
\citet{Elbaz2011AA...533A.119E} 			& Salpeter  	& FUV 			& FIR					& blue		    & PEGASE.2	& C00			& (0.7,0.3,0.7)		& GOALS/AKARI/SDSS   		& 38960   \\
\cite{Heinis:2014MNRAS.437.1268H}		& Chabrier	& UV+IR 			& UV						& None		    & BC03		& None			& (0.7,0.3,0.7)		& HerMES/COSMOS		& 	1.68 	\\
\citet{Magnelli2014AA...561A..86M} 	 		& Chabrier  	& UV+IR 			& $K$					& 3$\sigma$-clip   & BC03		& None			& (0.71,0.27,0.73)	& COSMOS				& 1.97      \\
\citet{Noeske:2007ApJ...660L..47N} 			& Kroupa    	& UV+IR 			& Optical 					& redshift		    & BC03		& $^g$CE01		& (0.7,0.3,0.7)		& AEGIS					& 1 		\\
\citet{Oliver2010MNRAS.405.2279O} 		& Salpeter  	& IR 				& FIR					& None		    & $^b$R08	& None			& (0.7,0.3,0.7)		& SWIRE					& 11.33    \\
\citet{Pannella2009ApJ...698L.116P}       		& Salpeter 	& 1.4 GHz 		& {\it sBzK}				& {\it sBzK}	    & BC03		&  P84			& (0.7,0.3,0.7)		& COSMOS				& 0.9        \\	
\citet{Reddy2012ApJ...754...25R}             		& Salpeter  	& UV+IR 			& UV						& LBG		    & $^c$CB11	& C00			& (0.7,0.3,0.7) 		& various					& 0.081    \\
\citet{Rodighiero2010}					& Salpeter		& UV+IR	 		& 4.5\,$\mu$m				& {\it U-B}		    & BC03		& C00			& (0.7,0.27,0.73)       & GOODS				&?		\\
\citet{Rodighiero2011ApJ...739L..40R}  		& Salpeter  	& UV+IR  			& NIR+FIR				& $sBzK$		    & BC03		& C00			& (0.73,0.26,0.74)	& COSMOS/GOODS-MUSIC	& 1.73 	\\ 
\citet{Santini2009AA...504..751S} 			& Salpeter 	& UV+IR			& $z$/$K$/4.5\,$\mu$m 		& 2$\sigma$-clip   & M05		& C97/C00		& (0.7,0.3,0.7)		& GOODS-MUSIC 			& 0.04	\\
\citet{Schreiber2015AA...575A..74S}     		& Salpeter 	& UV+IR	        		& $H$/$K$	 			& {\it UVJ}		    & BC03		& C00			& (0.7,0.3,0.7)		& CANDELS-GOODS/COSMOS & 1.8	\\
\citet{Whitaker2012} 						& Kroupa  	& UV+IR 			& NIR					& {\it UVJ}	             & BC03	& C00			& (0.7,0.3,0.7)		& NMBS					& 0.4		 \\	
\citet{Whitaker2014ApJ...795..104W} 		& Chabrier	& UV+IR/H$\beta$	& NIR					& {\it UVJ}		    & BC03		& C00			& (0.7,0.3,0.7)		& CANDELS/3D-HST		& 0.25       \\
\citet{Zahid2012ApJ...757...54Z}              		& Chabrier 	& H$\alpha$/H$\beta$ & Optical/UV				& Lines		    & BC03		& C00			& (0.7,0.3,0.7)	 	& SDSS-DR7/DEEP2		& 0.5 	\\	
\\
\multicolumn{6}{c}{$^a$\cite{fiocroccavolmerange97},$^b$\cite{Rowan2008MNRAS.386..697R}, $^c$Charlot \& Bruzual (2011) (never formally published)}\\
\multicolumn{5}{c}{$^d$\cite{calzetti00}, $^e$\cite{Prevot1984AA...132..389P},$^f$\cite{charlotfall00}, $^g$\cite{CE01}}\\
\hline
\end{tabular}
\end{table*}

To summarise, we have found that our dominant source of calibration required for our comparison is from the SPS models.  Our total calibration offsets in the normalisation of our `raw' MS fits  can be found on Table~\ref{sec:cal} (shown in parenthesis)  along with the S14 derived offsets for the works we compare to in Table~\ref{tab:assumps}.

\section{Results}\label{sec:results}

Figure~\ref{fig:vidmedzbest}  shows the SF-MS for our SFG sample across twelve redshift bins. The grey region in each panel shows the \mstar\ completeness limit;  galaxies within this region are excluded from our analysis. The blue and white circles show the median SFRs in bins of width  $\Delta M_*$ = 0.2 dex.  
 
We find a consistent  correlation between SFR and \mstar\ out to $z<3.0$ which we  quantify with the correlation coefficient, $r_c$, in each redshift panel on the Figure.  We find  an average value of $r_c~=~0.61$; a trend consistent with many other works \citep[e.g.][]{Erb2006ApJ...647..128E,daddi07,Noeske:2007ApJ...660L..47N,Dunne:2009MNRAS.394....3D,Pannella2009ApJ...698L.116P,Rodighiero2010,Elbaz2011AA...533A.119E,Whitaker2012,Muzzin2013ApJS..206....8M,Magnelli2014AA...561A..86M}.   We have also estimated the dispersion of the SF-MS denoted as \sigms\ in Figure~\ref{fig:sigmams}, for four stellar mass bins with $\Delta$\lgmstar\ = 0.5 dex, centred on \lgmstar = 10.0 (red), 10.5 (blue), 11.0 (green) and 11.5 (black). Previous studies by e.g.   \citet{Noeske:2007ApJ...660L..47N,Whitaker2012} and \cite{Magnelli2014AA...561A..86M} have reported an average dispersion over all stellar mass ranges between $\sim$0.3 and 0.35, which is indicated by the shaded grey region in Figure~\ref{fig:sigmams}.  As we can see, our values for \sigms\ vary about this region as a function of redshift, where in general at $z\lesssim 1.4$ we find $0.27<\sigma_{MS}<0.32$, at $1.4<z<1.6$ the scatter  increases to \sigms $\sim0.38$ before showing an overall decline towards $z\sim3$ for the highest stellar mass bins with \lgmstar=11.0 and 11.5. This decline in dispersion was also alluded to in \citet{Whitaker2012} and \citet{daddi07}.   If we take the average values of \sigms\ for each stellar mass bin across their respective redshift ranges we find for \lgmstar=10.0, 10.5, 11.0 and 11.5 dispersions of \sigmsmean = 0.29, 0.29, 0.31 and 0.28 respectively.

\begin{figure}
\centering
 \includegraphics[width=0.49\textwidth]{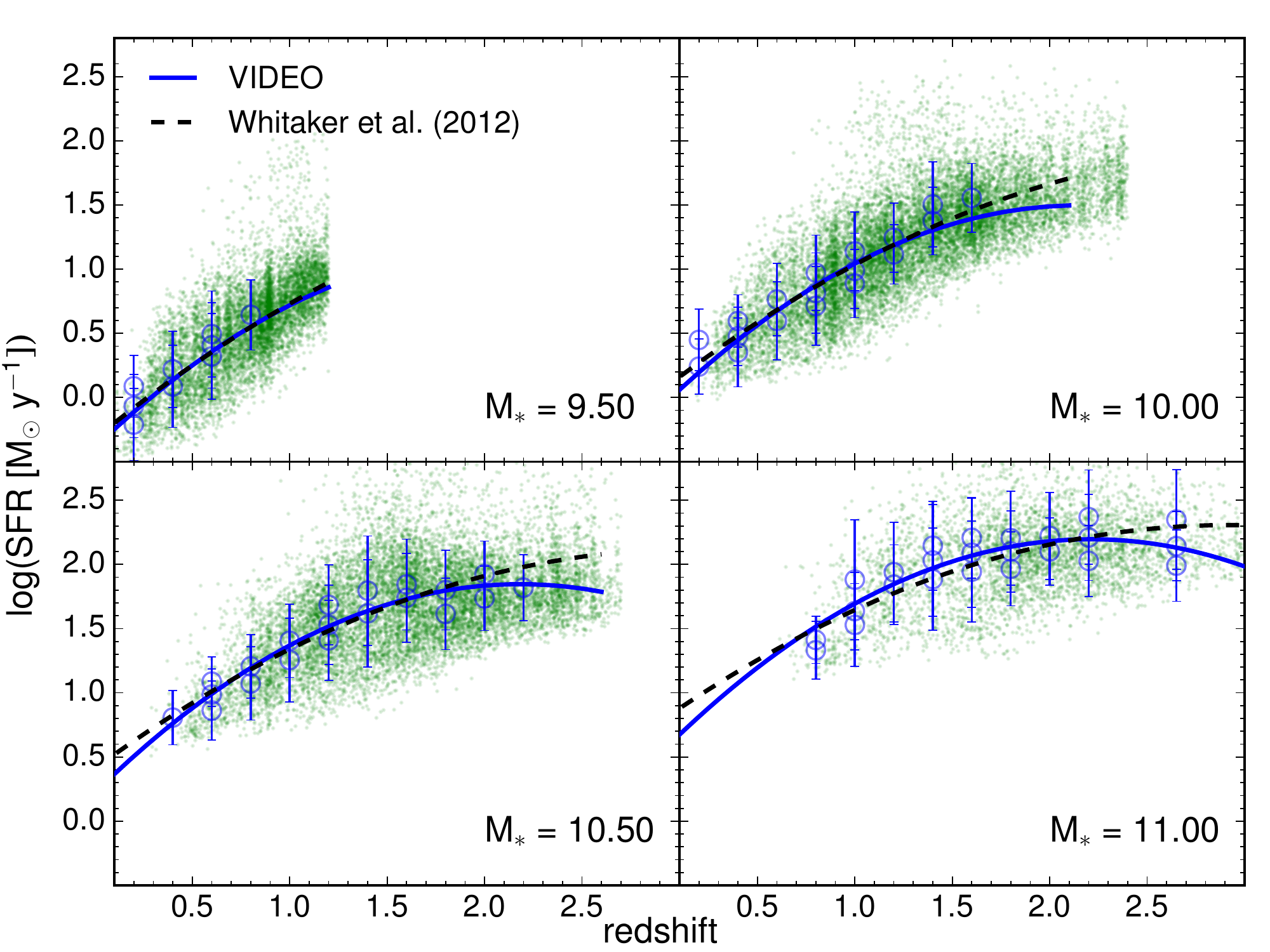}
     \caption{\small  SFR versus redshift for the SFGs in our sample, in stellar mass bins of \lgmstar\ = 9.5, 10.0, 10.5 and 11.0. In each panel the green points show the galaxy distribution, and the blue circles denote the median values used in the fit from Figure~\ref{fig:vidmedzbest}.  The redshift-dependent fit to the median points (Equation~\ref{equ:W12})  is shown as the blue solid line.  The black dashed line is the best fit from \citet{Whitaker2012} shown for comparison. }
    \label{fig:sfrzscatter}    
\end{figure} 

\begin{table*}
\caption{SFR-MS parameterised according to Equation~\ref{equ:W12}. The
  second column shows the results when we propagate the
  full redshift PDFs through \cig,  computed as the mean of the 1000
  Monte Carlos fits with their respective averaged \sig\ standard deviations. The third column is fitted to the full observed data-set including the intrinsic errors on stellar mass and SFR from \cig\ ($\sigma_{\rm cig}^2$), the uncertainty 
  on the best-fit redshift from \lephare\ ($\sigma_{\rm leph}^2$), the intrinsic dispersion of the SFR and stellar mass distributions ($\sigma_{\rm disp}^2$), and
  finally the uncertainty due to cosmic variance \sigv$^2$.
  The forth column shows our fits to the median values and the last
  column shows the fits to the median values from   
 \citet{Whitaker2012}.  The reduced $\chi^2/\nu$ is shown in the bottom row.}
\begin{tabular}{cccccc}
 		&  \multicolumn{3}{c}{This Work}& \citet{Whitaker2012} \\
                 &  \multicolumn{3}{c}{($0.1<z_{ph}\le3.0$)}   &  ($0<z<2.5$) \\
Parameter             &  \multicolumn{3}{c}{$\overbrace{\rule{7cm}{0pt}}$}  \\
(Equ. ~\ref{equ:W12})& $z_{ph}$-PDF (all data) & [$\sigma_{\rm cig}^2$+$\sigma_{\rm leph}^2$+$\sigma_{\rm disp}^2$+\sigv$^2$]     &  (medians)  &(medians)\\ 
\hline\hline
$\alpha_1$  &   0.68   $\pm$ 0.01	&  0.68   $\pm$ 0.01      &   0.62    $\pm$ 0.04    &  0.73 $\pm$ 0.16  \\  
$\alpha_2$  &	0.004 $\pm$ 0.011    &  0.004 $\pm$ 0.011   &	0.03 $\pm$ 0.04     &  -0.12 $\pm$ 0.16  \\
$\beta_1$    &	0.33   $\pm$ 0.02    	&  0.29   $\pm$  0.02    &	0.24 $\pm$ 0.05     & 0.41 $\pm$ 0.17   \\  
$\beta_2$    &	1.36   $\pm$ 0.03	&  1.45   $\pm$ 0.03  	& 	1.47 $\pm$ 0.08      &  1.11 $\pm$ 0.28  \\  
$\beta_3$    & 	-0.30  $\pm$ 0.01   	& -0.32   $\pm$ 0.01  & 	-0.33 $\pm$ 0.03   & -0.18 $\pm$ 0.11  \\\\ 
$\chi^2/\nu$ & 0.70   & 0.78  & 0.49  & 0.05 \\
\hline\hline
\end{tabular}
  \label{tab:polyfit}    
\end{table*}

\subsection{Main Sequence  constraints - I}

In this section we model the SF-MS relation using the redshift-dependent model given by  Equation~\ref{equ:W12} and to check the robustness of our results we model our data  with three approaches. In the first we explore the contribution from our \zph-PDF derived values for SFR and \mstar\ as detailed in \S~\ref{sec:zpdfcigale}. We then model our observed full sample that includes intrinsic errors from \cig\, and finally fit to  the median values of Figure~\ref{fig:vidmedzbest} which allows a more direct comparison to the work of \whit. Table~\ref{tab:polyfit} summarises the results of these  fits. 

In the second column in Table~\ref{tab:polyfit} we account for each \lephare\ \zph-PDF  by  creating a set of 1,000 Monte Carlo (MC) data-sets by  randomly  sampling from the PDFs and generating corresponding SFRs and stellar masses as described in \S~\ref{sec:zpdfcigale}. We then fit Equation~\ref{equ:W12}  to each MC sample where the likelihood term includes an additional uncertainty for the intrinsic dispersions of SFR-$z$ and \mstar-$z$ distributions as well as the contribution from cosmic variance \sigv.   In this way we can directly inspect the distribution of constrained parameters and estimate their  confidence limits.  Here, the marginalised  parameter constraints  are the mean of the 1000 MC sample distributions and the associated errors are the mean \sig\ standard deviations. In the third column of the table we fit Equation~\ref{equ:W12} to our actual observed sample to incorporate statistical uncertainties from the intrinsic errors from \cig\ on SFR and \mstar\, the uncertainty   on the best-fit redshift from \lephare, the intrinsic dispersion of the SFR-$z$ and \mstar-$z$ distributions, and the uncertainty due to cosmic variance. Our median results are then shown in the fourth column with the \whit\ constraints presented in the last column of the table. It should be noted that we have reanalysed their median data points using the MCMC routine to provide error estimates on the parameters that were not previously available.

In the left hand panel of Figure~\ref{fig:mcmc} we show an example of the 2-D 1$\sigma$ and 2$\sigma$ marginalised results of one of the MC  \zph-PDF fits compared to our full data set and median constraints. We find that they are consistent to within 2$\sigma$. 
On closer inspection of the \zph-PDFs outputted from \lephare\ we found only a small fraction of objects in our final sample were poorly constrained and therefore their contribution to this fit would have minimal impact. Nevertheless, this remains an important quality check of our sample to the robustness of the final result. Furthermore, comparing our median results to those of \whit\  (right panel set of  Figure~\ref{fig:mcmc}) and Figure~\ref{fig:sfrzscatter}, we find our results are  consistent to within their \sig\ errors, although we constrain the parameters much more strongly due to the larger sample.

\subsection{Main Sequence  constraints - II}\label{sec:plfits}

\begin{figure*}
\centering
 \includegraphics[width=1.0\textwidth]{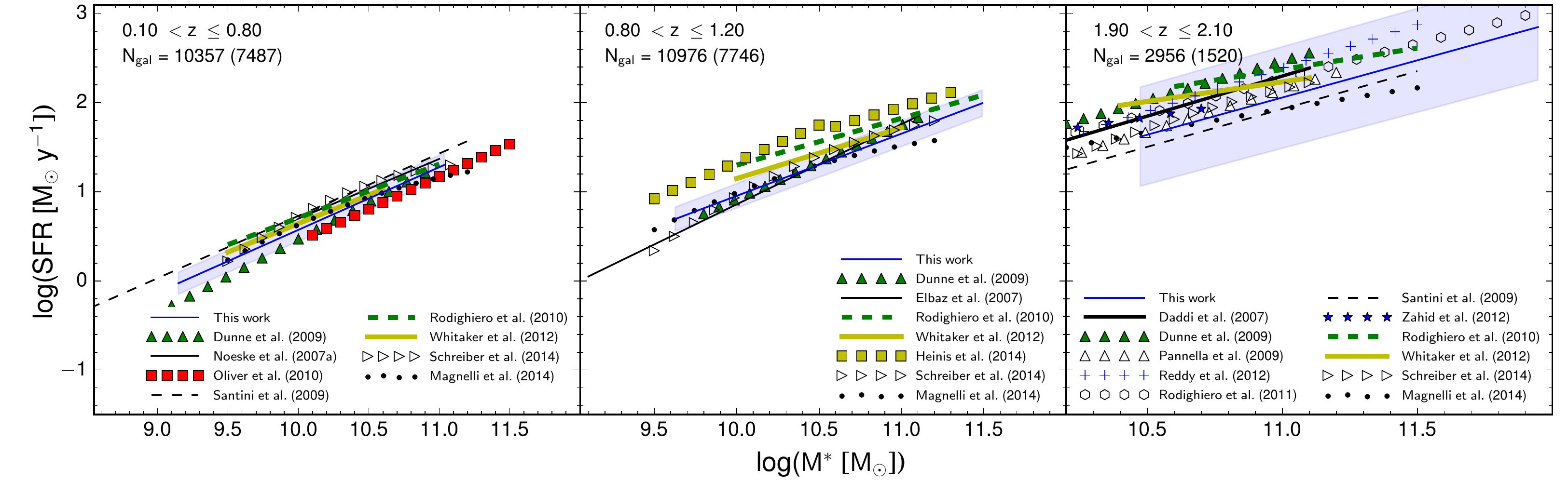}
     \caption{\small  Comparison of the SF-MS.  Our fits to the data are shown by the thin blue line with the error shown as the light blue shaded region.  For each panel we find a slope $\alpha= 0.70\pm0.01$, $0.70\pm0.01$ and $0.83\pm0.02$ at $z\sim 0.45$, 1 and 2 respectively.  Where possible we have shown comparative measurements within their respective stellar mass ranges from   \citet{Noeske:2007ApJ...660L..47N,daddi07,Elbaz2007AA...468...33E,Santini2009AA...504..751S,Dunne:2009MNRAS.394....3D,Pannella2009ApJ...698L.116P,Rodighiero2010,Oliver2010MNRAS.405.2279O,Elbaz2011AA...533A.119E,Rodighiero2011ApJ...739L..40R,Zahid2012ApJ...757...54Z,Reddy2012ApJ...754...25R, Whitaker2012, Heinis:2014MNRAS.437.1268H,Magnelli2014AA...561A..86M} and \citet{Schreiber2015AA...575A..74S}.  It should be noted that the symbols are not data points but merely indicate the power-law fit. The fits by \citet{Heinis:2014MNRAS.437.1268H} in the middle panel were estimated at $z\sim1.4$ using a broken power-law (see text for details).}
    \label{fig:sfrmasscomp}    
\end{figure*}

Table~\ref{tab:pwerlaw} shows our resulting power-law constraints as applied to the redshift bins in Figure~\ref{fig:vidmedzbest}. The term `all-data' refers to our fits within each redshift slice to the full mass-complete galaxy distribution and the values in parenthesis are our re-calibrated MS fits.  Firstly we note that the median and `all-data' fits are fully consistent within \sig\ and, as also found in the previous section, we observe strong evolution in the normalisation ($\beta$) consistent with the picture of an overall increase in star formation rate, at a given stellar mass, with redshift. In particular we observe a steady decline in SFR for a fixed stellar mass of   \lgmstar$\sim$10.0  by a factor of $\sim19$ from $z\sim2$ to $z\sim0.2$.  In the following sections we compare our results to the broader literature that coincide  at three selected redshift ranges $0.1<z<0.8$, $0.8<z<1.2$  and $1.9<z<2.1$. We use the calibrated MS fits as discussed in \S~\ref{sec:cal} which are shown in  Table~\ref{tab:fig11fits} and  Figure~\ref{fig:sfrmasscomp}. In all three panels our results are shown as the blue line with the error on the fits shown as a shaded blue region.

\subsubsection{Comparison at $z\sim0.45$}
\begin{figure}
\centering
 \includegraphics[width=0.49\textwidth]{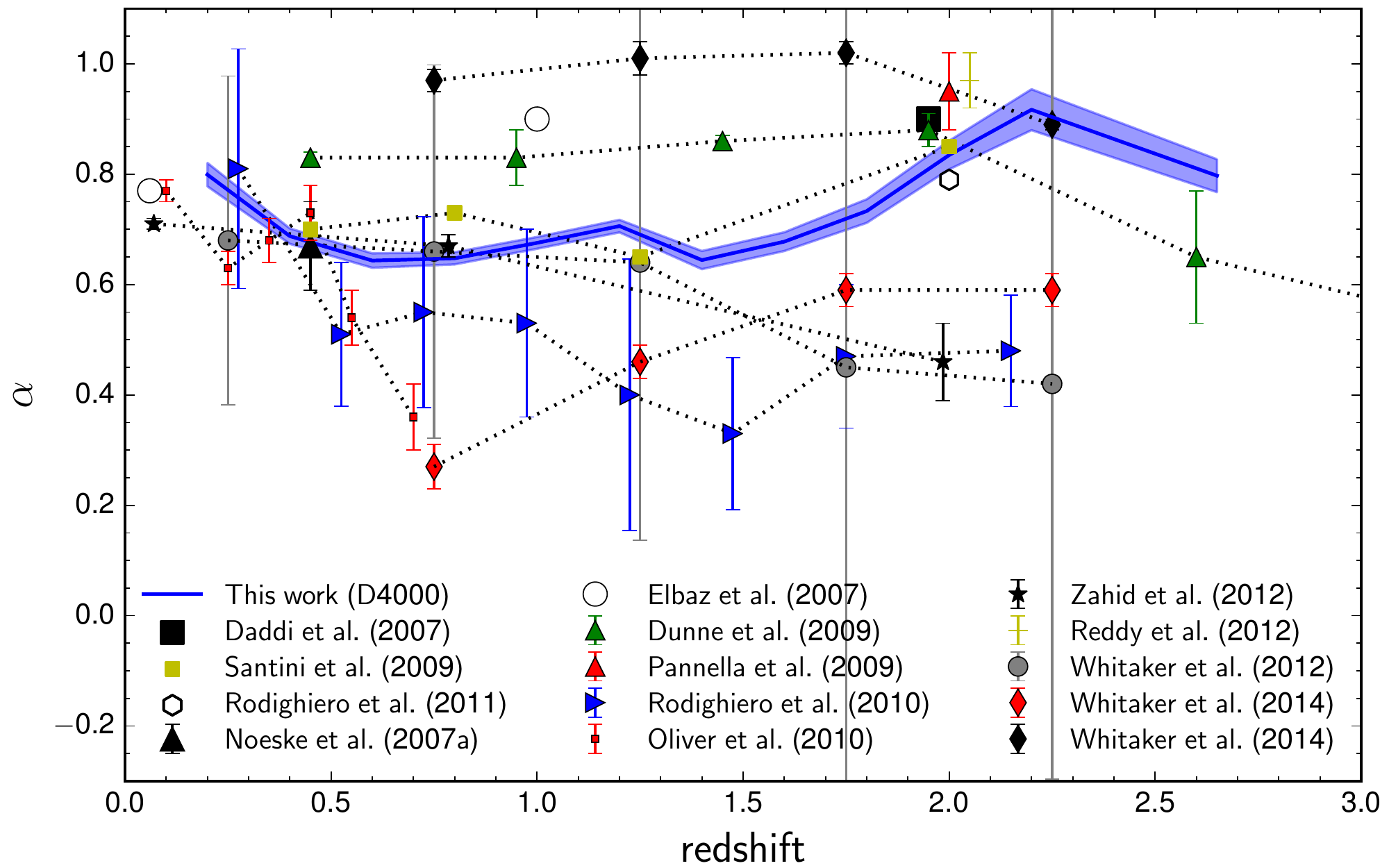}\hfill
     \caption{\small  Comparison of the power-law  slope $\alpha$ constraints  as a function of redshift.  The blue line and shaded error region is our fiducial result when we apply our $D4000$ selection.   The legend indicates comparative works from  \citet{Noeske:2007ApJ...660L..47N,daddi07,Elbaz2007AA...468...33E,Santini2009AA...504..751S,Dunne:2009MNRAS.394....3D,Pannella2009ApJ...698L.116P,Rodighiero2010,Oliver2010MNRAS.405.2279O,Elbaz2011AA...533A.119E,Rodighiero2011ApJ...739L..40R,Zahid2012ApJ...757...54Z,Reddy2012ApJ...754...25R}, \citet{Whitaker2012}  and \citet{Whitaker2014ApJ...795..104W} as also compared in Figure~\ref{fig:sfrmasscomp}. The \citet{Whitaker2014ApJ...795..104W} data represent fits for log stellar masses $< 10.2$ (solid red diamonds) and $>10.2$ (solid black diamonds).}
    \label{fig:betaz}    
\end{figure}

In  Figure~\ref{fig:sfrmasscomp} we show our fits to the `all-data' case (denoted by the blue shaded region).    In the left-hand  panel of the figure our data covers a redshift range of  $0.10<z\leq 0.80$ with  a  stellar mass complete sample of 7487 galaxies.
At this redshift we find  [$\alpha$, $\beta$] = [$0.70 \pm 0.01$,$-6.46 \pm 0.06$], which are fully consistent with \cite{Noeske2007ApJ...660L..43N} who found $[ 0.67 \pm 0.08,-6.00 \pm 0.78]$ (black solid line),   \cite{Santini2009AA...504..751S}  $[0.70\pm 0.14,-6.27 \pm1.54]$  (dashed black line) and \cite{Oliver2010MNRAS.405.2279O} $[0.73 \pm 0.05,-6.86 \pm 0.55]$ (red squares). The radio stacking results by  \cite{Dunne:2009MNRAS.394....3D} find a much steeper gradient to ours with constraints  $[\alpha, \beta]~=~ [0.83~\pm~0.01,-7.83~\pm~0.13]$. In the case of \cite{Whitaker2012}, their data was fitted using the redshift-dependant polynomial model of Equation~\ref{equ:W12}, which is represented by the yellow solid line. Since their redshift binning is not aligned with ours we cannot make a straightforward power-law fit comparison, however, as we can see in the left panel of Figure~\ref{fig:sfrmasscomp}, their fit seems very closely aligned with that of \cite{Noeske2007ApJ...660L..43N} in both normalisation and slope over this redshift range. In Figure~\ref{fig:betaz} we show the power-law fit to their data for each redshift bin (shown as black solid circles) and we find good agreement with our result over this redshift range. 

\begin{table}
\caption{Power-law contraints from Figure~\ref{fig:sfrmasscomp}. The numbers in parenthesis are those calibrated
by \citet{Speagle2014ApJS..214...15S} and which were used in the plotting of the Figure. Details of our calibration can be
found in \S~\ref{sec:cal}.}

\label{tab:fig11fits}
\centering
\begin{tabular}{l c c }

Redshift     &  $\alpha$ & $\beta$   \\

\hline\hline
$0.1<z<0.8$ \\ 
This work							& 0.70 $\pm$ 0.01    &  -6.36 (-6.46) $\pm$ 0.06 \\
\citet{Dunne:2009MNRAS.394....3D}     	& 0.83 $\pm$ 0.01  	& -7.92 (-7.83) $\pm$ 0.13  \\		
\citet{Noeske:2007ApJ...660L..47N} 		& 0.67 $\pm$ 0.08 	& -6.19 (-6.00) $\pm$ 0.78 \\		
\citet{Oliver2010MNRAS.405.2279O} 	& 0.73 $\pm$ 0.05 	& -7.02 (-6.86) $\pm$ 0.55  \\		
\citet{Santini2009AA...504..751S} 		& 0.70 $\pm$ 0.14 	& -6.33 (-6.27) $\pm$ 1.54  \\\\		

$0.8<z<1.2$ \\ 
This work							& 0.70 $\pm$ 0.01   & -5.92 (-6.02)  $\pm$ 0.07 \\
\citet{Dunne:2009MNRAS.394....3D}         & 0.83 $\pm$ 0.05 	& -7.38 (-7.39) $\pm$ 0.53 \\ 
\citet{Elbaz2007AA...468...33E} 		& 0.90 			& -8.14 (-8.06)			 \\\\

$1.9<z<2.1$ \\ 
This work							 &  0.84 $\pm$ 0.04   &-7.01  (-7.13)  $\pm$ 0.28 \\
\citet{daddi07}  						 & 0.90 			& -7.60 (-7.52)			\\
\citet{Dunne:2009MNRAS.394....3D}          & 0.88 $\pm$ 0.03  	& -7.21(-7.36)    $\pm$  0.37 \\
\citet{Pannella2009ApJ...698L.116P}          & 0.95 $\pm$ 0.07	& -8.30 (-8.35)			 \\	
\citet{Reddy2012ApJ...754...25R}               & 0.97 $\pm$ 0.05   & -8.28  (-8.32)  $\pm$ 1.28    \\
\citet{Rodighiero2011ApJ...739L..40R}  	 & 0.79 			& -6.42 (-6.36)			 \\
\citet{Santini2009AA...504..751S}       	& 0.85 $\pm$ 0.17  & -7.24 (-7.15)    $\pm$ 1.87 \\
\citet{Zahid2012ApJ...757...54Z}             	& 0.46  $\pm$ 0.07  &  -2.99 (-2.94)  $\pm$ 0.70 \\\\

\hline
\end{tabular}
\end{table}

In a similar way, \citet{Magnelli2014AA...561A..86M} modelled their data with a quadratic given by Equation~\ref{equ:magnelli}. In this case we  have averaged their results over their two redshift ranges $0.20<z\leq 0.50$ and $0.50<z\leq 0.80$ to obtain an approximate comparison. Whilst their data supports a turnover toward the higher stellar masses, there appears to be a good agreement up to \lgmstar$\sim$10.5 at which point their fit diverges. Although we note that no errors on their parameter constraints were reported in their analysis. \cite{Schreiber2015AA...575A..74S} also model their data in such a way as to allow for a SFR turnover (Equation~\ref{equ:schreiber}) which we can observe at this redshift similar to that of  \citet{Magnelli2014AA...561A..86M}.

As with  \cite{Whitaker2012}, the binning of \cite{Rodighiero2010} are also not aligned with ours. However, we show our fits to their data in Figure~\ref{fig:betaz}, and for illustrative purposes have interpolated their fits to show their results in Figure~\ref{fig:sfrmasscomp}. As a cross-check we fitted their data to the redshift-dependent model Equation~\ref{equ:W12} and found these fits to be consistent with our interpolated results. At this redshift their results show a slightly shallower slope of $\alpha=0.60$ and normalisation $\alpha=-5.38$.

It is perhaps not too surprising that there is, in general, very good agreement between the results shown here at this relatively low redshift, since there is less likely to be residual contamination from passive galaxies which could  result in shallower slopes. 

\begin{table*}
\caption{Parameter constraints for $\log_{10}({\rm SFR}) =  \alpha\log_{10} (M_*)+\beta$. The redshift slices correspond to those in Figure~\ref{fig:vidmedzbest}. Columns 2 and 3  the fits to the median values of SFR. Columns 4 and 5 are  fitted to the full distribution of data points. The values in parenthesis are our re-calibrated fits from M05 to BC03 SPS models (see \S~\ref{sec:cal} for details).}
\begin{center}
\begin{tabular}{cccccc}
&  \multicolumn{2}{c}{median fit} &  \multicolumn{2}{c}{all data fit} \\
&     \multicolumn{2}{c}{$\overbrace{\rule{4cm}{0pt}}$} &  \multicolumn{2}{c}{$\overbrace{\rule{4cm}{0pt}}$} \\\
Redshift slice & $\beta$ & $\alpha$ & $\beta$  & $\alpha$  \\

\hline\hline
$0.10 <z\leq 0.30$ & $-7.86\ (-7.98)\ \pm 2.30$  & $0.82 \pm 0.24$ & $-7.64\ (-7.76)\ \pm 0.20$ & $0.80 \pm 0.02$ \\
$0.30 <z\leq 0.50$ & $-6.27\ (-6.37)\ \pm 2.18$  & $0.68 \pm 0.22$ & $-6.35\ (-6.45)\ \pm 0.13$ & $0.69 \pm 0.01$ \\
$0.50 <z\leq 0.70$ & $-5.65\ (-5.74)\ \pm 2.35$  & $0.63 \pm 0.23$ & $-5.73\ (-5.82)\ \pm 0.13$ & $0.64 \pm 0.01$ \\
$0.70 <z\leq 0.90$ & $-4.79\ (-4.88)\ \pm 2.05$  & $0.56 \pm 0.20$ & $-5.60\ (-5.69)\ \pm 0.11$ & $0.65 \pm 0.01$ \\
$0.90 <z\leq 1.10$ & $-6.10\ (-6.21)\ \pm 2.23$  & $0.71 \pm 0.21$ & $-5.69\ (-5.79)\ \pm 0.11$ & $0.68 \pm 0.01$ \\
$1.10 <z\leq 1.30$ & $-5.75\ (-5.86)\ \pm 2.59$  & $0.69 \pm 0.25$ & $-5.86\ (-5.96)\ \pm 0.12$ & $0.71 \pm 0.01$ \\
$1.30 <z\leq 1.50$ & $-4.94\ (-5.03)\ \pm 3.49$  & $0.63 \pm 0.33$ & $-5.07\ (-5.16)\ \pm 0.17$ & $0.64 \pm 0.02$ \\
$1.50 <z\leq 1.70$ & $-4.94\ (-5.03)\ \pm 3.79$  & $0.64 \pm 0.35$ & $-5.35\ (-5.45)\ \pm 0.18$ & $0.68 \pm 0.02$ \\
$1.70 <z\leq 1.90$ & $-5.47\ (-5.57)\ \pm 4.26$  & $0.69 \pm 0.39$ & $-5.93\ (-6.04)\ \pm 0.23$ & $0.73 \pm 0.02$ \\
$1.90 <z\leq 2.10$ & $-5.93\ (-6.04)\ \pm 4.53$  & $0.73 \pm 0.41$ & $-7.00\ (-7.12)\ \pm 0.28$ & $0.83 \pm 0.03$ \\
$2.10 <z\leq 2.30$ & $-7.36\ (-7.49)\ \pm 5.30$  & $0.87 \pm 0.48$ & $-7.86\ (-8.00)\ \pm 0.40$ & $0.92 \pm 0.04$ \\
$2.30 <z\leq 3.00$ & $-7.44\ (-7.58)\ \pm 4.94$  & $0.88 \pm 0.44$ & $-6.54\ (-6.66)\ \pm 0.33$ & $0.80 \pm 0.03$ \\
\hline\hline

\end{tabular} \label{tab:pwerlaw}
\end{center}
\end{table*}

\subsubsection{Comparison at $z\sim1$}
As we move to the intermediate redshift bin $0.80<z\leq 1.2$ shown in the middle panel of Figure~\ref{fig:sfrmasscomp}, we have a total of 7746 galaxies  in our stellar mass complete sample.   Here we find very little difference from our result at $z\sim0.45$ where  $[\alpha, \beta] =[0.70 \pm 0.01,-6.02 \pm 0.07]$.  However, we now begin to see more divergent trend  between data-sets. For example,  \cite{Elbaz2007AA...468...33E}  (black solid line)  found a slope $\alpha =0.9$, close to unity and a significantly lower normalisation of $\beta=-8.14$. A similar trend was found by   \cite{Dunne:2009MNRAS.394....3D} with  $[\alpha, \beta] = [0.83 \pm0.05,-7.39 \pm 0.53]$  (green triangles). In contrast,  \cite{Rodighiero2010} find a shallowing slope of $\alpha$ to 0.52. At this redshift our results are more aligned with  \cite{Whitaker2012} and \citet{Magnelli2014AA...561A..86M}. We also note that  \cite{Santini2009AA...504..751S}  found a slope that becomes shallower between $0.6<z<1.5$ from $\alpha = 0.73$ at $z\sim0.8$ to $\alpha=0.65$ at $z\sim1.2$.  With our D4000  SFG  selection criteria we do not find strong evidence for a turnover in the SF-MS relation, however, we find that over this redshift range our results are consistent with those of  \citet{Magnelli2014AA...561A..86M} (black dotted line) and \cite{Schreiber2015AA...575A..74S} (right triangles), at least up to \mstar$\sim$10.5. 
The results shown by \cite{Heinis:2014MNRAS.437.1268H} are consistent with ours with a modelled power-law giving [$-5.7\pm0.7,0.70\pm0.07$]. However, they found a broken power-law a better fit to their data at this redshift finding  a slope of  $\sim$0.85 for \mstar $< 10^{10.5} M_\odot$   and a turning over to  $\sim$0.5 at the high mass end (shown as yellow squares in the Figure).  In Section~\ref{sec:sfsel} we will explore the sensitivity of our modelling with respect to how SFGs are selected.

\subsubsection{Comparison at $z\sim2$}
Finally, in the third panel we show the SF-MS over the redshift range $1.90<z\leq2.10$, with 1683 galaxies considered in the fit. Here we find significant positive evolution of the slope towards unity with   [$\alpha$, $\beta$] =[$0.84 \pm 0.04$,$-7.13  \pm 0.28$].   Here we find  separations in the magnitude of the slopes. In general   \cite{Rodighiero2010}, \cite{Whitaker2012}, \cite{Zahid2012ApJ...757...54Z} and  \citet{Magnelli2014AA...561A..86M} favour a much shallower slope of $\alpha\sim0.45$, whereas we, along with    \cite{daddi07,Dunne:2009MNRAS.394....3D, Santini2009AA...504..751S,Rodighiero2011ApJ...739L..40R,Pannella2009ApJ...698L.116P} and \cite{Reddy2012ApJ...754...25R}, find the opposite, with a range $0.79\lesssim\alpha\lesssim0.97$. \cite{Schreiber2015AA...575A..74S} also reported a steeper relation towards higher redshifts with a turnover only observed in the lower redshift regime. It should be noted that the uncertainties on the parameters for the power-law fits within each redshift bin for the \cite{Whitaker2012} data are large, of the order $\sigma_\alpha \sim0.75$, which means that their results are formally consistent with ours at this redshift. The differences in $\alpha$ are perhaps better seen in Figure~\ref{fig:betaz}, which shows that the magnitude of $\alpha$ in our sample clearly begins to rise at $z\gtrsim1.5$ and e.g.  \cite{Whitaker2012} begins to fall.

\subsection{Evolution of specific star-formation rate}\label{sec:ssfr}
\begin{figure}
\centering
 \includegraphics[width=0.48\textwidth]{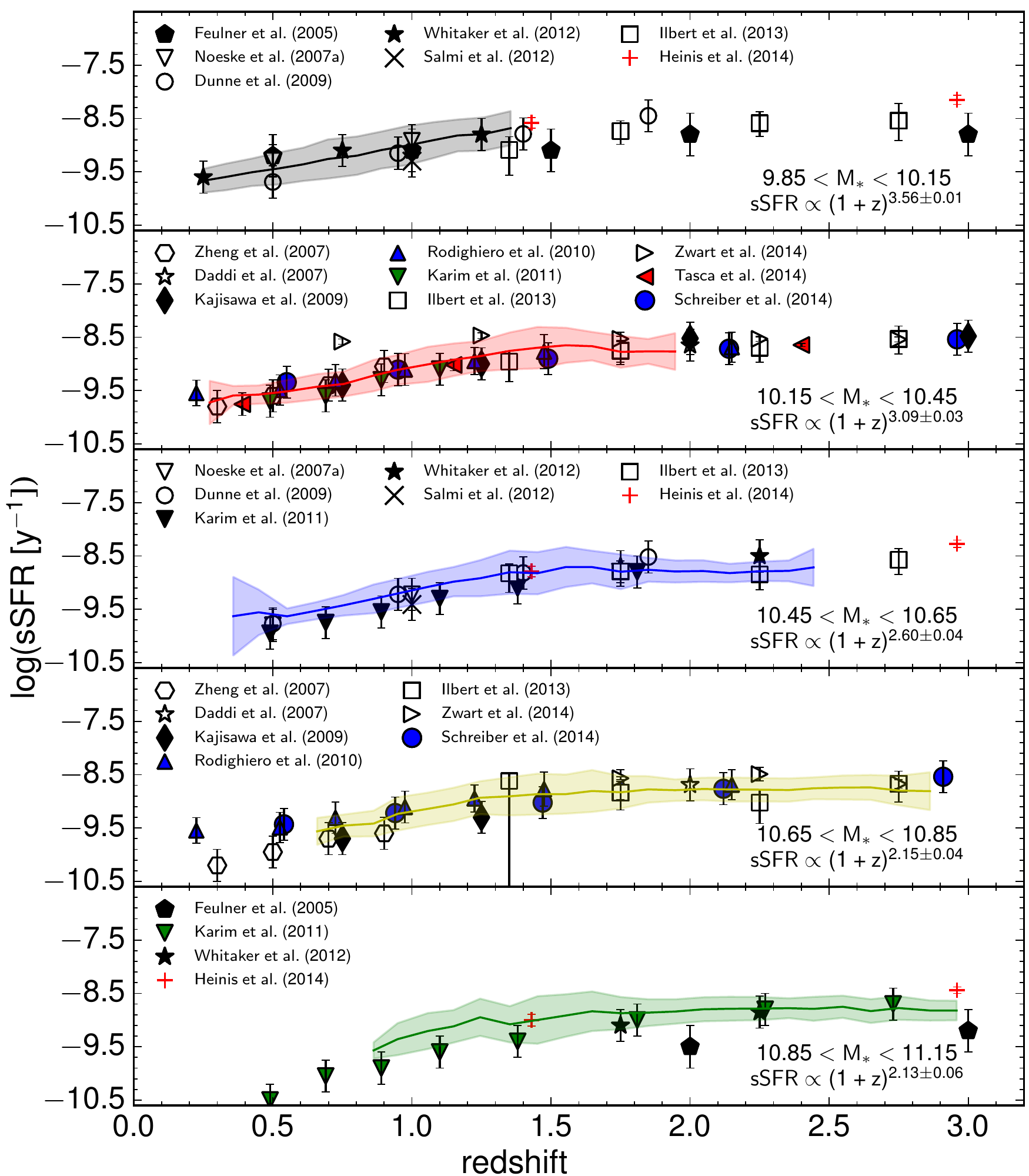}
     \caption{\small   SSFR evolution. On each panel we show our median values of SSFR as a function of redshift for a range of log stellar masses centred on \lgmstar=10.0,10.3,10.55,10.75 and 11.0, shown as black, red, blue, yellow and green shaded regions respectively. The statistical uncertainty is the combined error on the median and the cosmic variance  i.e.  \errtot=$\sqrt{(\sigma_{\rm MAD}^2+\sigma_{\rm V}^2)}$.  The other sSFR measurements are from  \citet{Feulner2005ApJ...633L...9F,Noeske:2007ApJ...660L..47N,daddi07,Dunne:2009MNRAS.394....3D,Kajisawa2009ApJ...702.1393K,Rodighiero2010,Karim2011ApJ...730...61K,Salmi2012ApJ...754L..14S,Whitaker2012,Ilbert:2013AA...556A..55I,Zwart2014MNRAS.439.1459Z,Tasca2014arXiv1411.5687T}, and \citet{Schreiber2015AA...575A..74S}.
     }\label{fig:ssfr-z}
\end{figure}

An alternative way to probe star formation histories over cosmic time is to look instead at the specific star formation rate (sSFR). This is simply 
given by sSFR=SFR/\mstar\ which allows us to probe star formation per unit stellar mass. In Figure~\ref{fig:ssfr-z} we show how the sSFR evolves as a function of redshift for our sample of SFGs for stellar-mass bins, centred on \lgmstar = 10.0, 10.3, 10.5, 10.75 and 11.0.  The redshift ranges we probe within each mass bin are consistent with our mass completeness limits.
We again compare our findings with other available sSFR measurements from the literature, as indicated in Figure~\ref{fig:ssfr-z}.
Our results are entirely consistent with other work out to $z\sim 3$ within our stellar-mass completeness limits, with our data showing a flattening off in the evolution of the sSFR beyond $z>1.5$ for our highest stellar-mass bins.

\begin{table}
\caption{Parameter constraints for (sSFR / yr$^{-1}$) = $A+ \gamma\log_{10}(1+z)$ for a range of fixed stellar masses as shown in Figure~\ref{fig:ssfr-z}.
}
\centering
\scriptsize
\begin{tabular}{cccc}
Stellar mass range & Redshift range & A & $\gamma$  \\
\lgmstar \\

\hline
$9.85<M_*<10.15$    & $0.25 <z\leq 1.35$ & $-10.05  \pm 0.01$  	& $3.56 \pm 0.01$ \\
$10.15<M_*<10.45$  & $0.27 <z\leq 1.95$ & $-10.05 \pm 0.01$	& $3.09\pm 0.03$ \\
$10.45<M_*<10.65$  & $0.36 <z\leq 2.44$ & $-10.03 \pm 0.01$  	& $2.60\pm 0.04$ \\
$10.65<M_*<10.85$  & $0.65 <z\leq 2.86$ & $-9.88   \pm 0.01$  	& $2.15 \pm 0.04$ \\
$10.85<M_*<11.15$  & $0.86 <z\leq 2.96$ & $-9.94   \pm 0.01 $  	& $2.13 \pm 0.06$\\

\hline\hline

\end{tabular} \label{tab:ssfr}
\end{table}

We parameterise the redshift evolution of the sSFR  for each stellar mass bin  with (sSFR / yr$^{-1}$) = $A+ \gamma\log_{10}(1+z)$.  These results are shown in Table~\ref{tab:ssfr}. We find that between mass bins 10.3 and 10.5 the relation evolves from $\gamma=3.09$ to 2.60  out to a redshift of 1.95 and 2.44, respectively. This result is consistent with that of the recent work from \cite{Tasca2014arXiv1411.5687T} who found $\gamma=2.8\pm0.2$ over a similar  redshift range for fixed stellar mass of  \lgmstar$\sim10.3$  using VUDS. 
Overall, we find a general flattening of the relation with redshift as we move to our highest  stellar mass bin of \lgmstar=11 with $\gamma=2.13\pm0.06$  and normalisation of $A=-9.94\pm0.01$ between $0.85<z<2.96$. 

If we restrict the redshift range out to $z<1.4$, following \cite{Ilbert2014arXiv1410.4875I},  
we find a similar, but slightly steeper evolution with $\gamma$=[3.39$\pm$0.01, 3.57$\pm$0.04,  3.39$\pm$0.05, 3.58$\pm$0.22, 3.91$\pm$0.36] for respective mass bins of \lgmstar = 9.5-10.0, 10-10.5, 10.5-11 and 11-11.5.  In their work they found a similar mass-dependent evolution in sSFR with $\gamma$=[2.88$\pm$0.12, 3.31$\pm$0.10,  3.52$\pm$0.15, 3.78$\pm$0.60]. Why  we find steeper slope at \lgmstar = 10 remains unclear.  However, we do note that at this mass range our fit was limited to $z\lesssim1.35$ due to our completeness limits. If we relax our SFG D4000 selection cut from e.g. 1.3 to 1.35, which would introduce an older population of galaxies into our sample, we find that this has only marginal affect. Alternatively, if we instead use a \ur\ colour selection as discussed in \S~\ref{sec:sfselect}, this only really affects the high mass bins which is consistent with our  findings in  \S~\ref{sec:sfsel}. Of course there remains the possibility of numerous options within \cig\ that could be explored in more detail, but we reserve this for future work. 

Nevertheless, within the redshift range we are considering in this work, we find very good agreement  across a broad range of related works, that shows an in increasing mass-dependent evolution with stellar mass below $z\sim2$, which then begins to slow considerably beyond this redshift.

\subsection{Star-forming selection effects and the high-mass turn over}\label{sec:sfsel}
\begin{figure*}
\centering
 \includegraphics[width=0.485\textwidth]{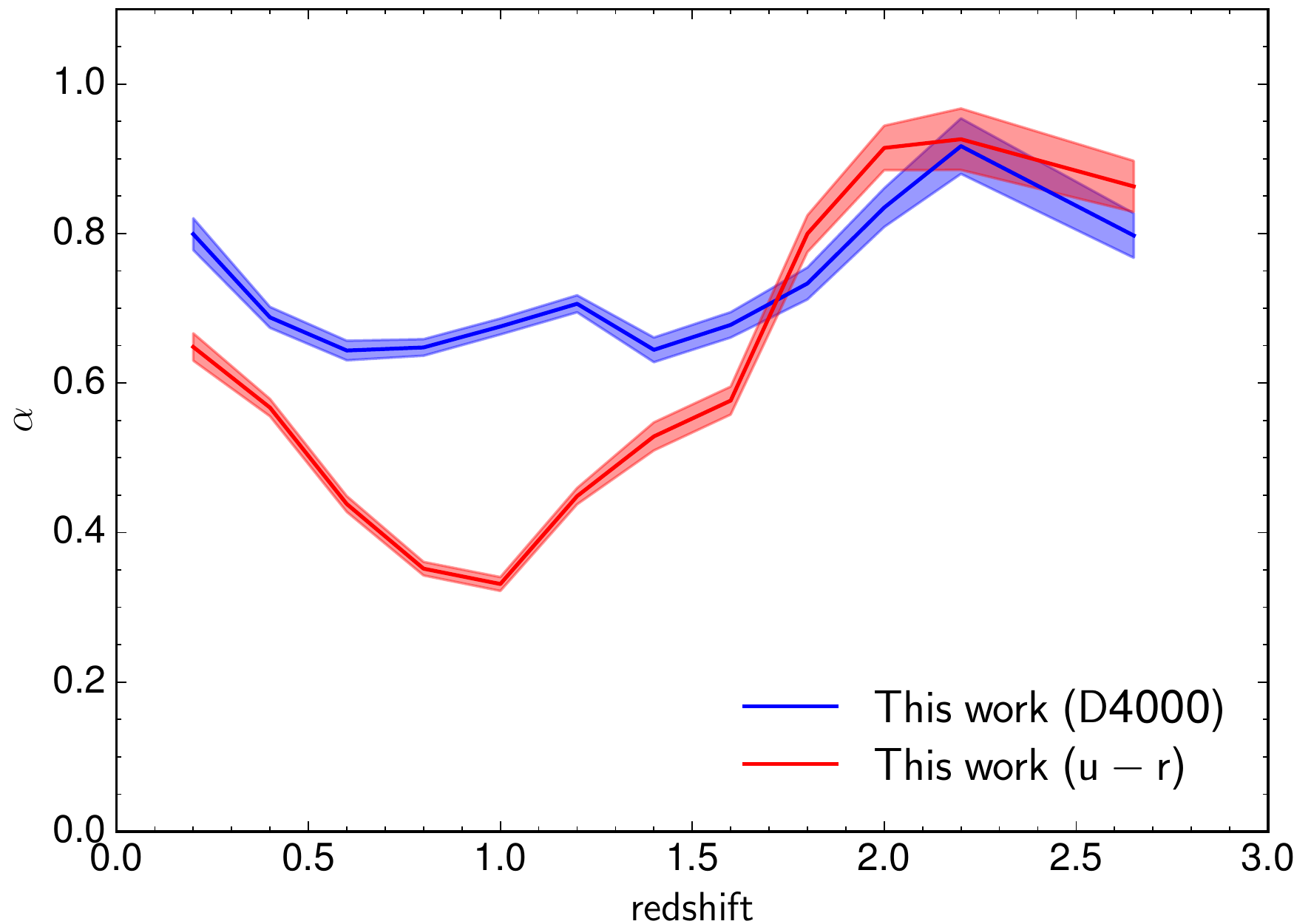}\hfill
  \includegraphics[width=0.49\textwidth]{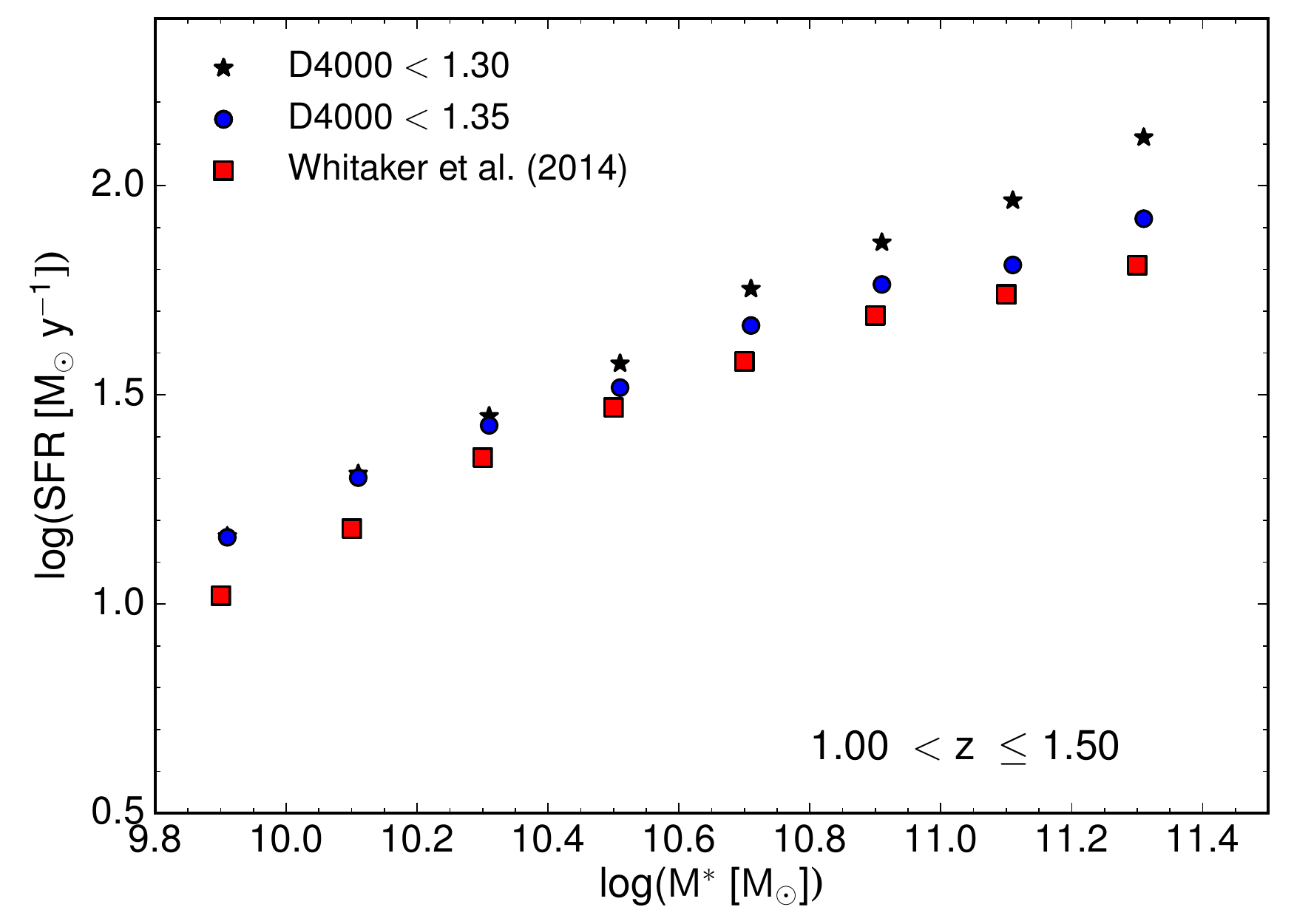}
     \caption{\small   {\it Left:} Comparison of derived power-law slopes ($\alpha$) as a function of redshift for our different SFG selection. The blue is for the D4000 < 1.3 cut  as was shown in Figure~\ref{fig:betaz} and the red shows the results for a rest-frame \ur\ selection.  {\it Right:} Comparing the high-mass turnover. The black stars show our fiducial result using an SFG criterion of D4000 $<1.30$. The blue circles show how our results change if we allow a slightly higher cut at 1.35, allowing older galaxies to be including in the SF-MS. We compare this to \citet{Whitaker2014ApJ...795..104W} data shown as red squares which presented results with flattening of the SF-MS \lgmstar\ $> 10.2$ dex. }
    \label{fig:d4sel}    
\end{figure*}

 As we discussed in \S~\ref{sec:cal}, the majority of off-sets added to calibrate the MS relation changes only  the normalisation.  However, the slope of the SF-MS is sensitive to effects such as how SFGs are selected \citep[e.g.][]{Oliver2010MNRAS.405.2279O,Whitaker2012,Rodighiero2014MNRAS.443...19R,Speagle2014ApJS..214...15S,Sobral2014MNRAS.437.3516S,Ilbert2014arXiv1410.4875I}.  As shown in Figure~\ref{fig:betaz} there is a broad range in the constrained values of  $\alpha$ as a function of redshift across the literature.  In the left panel of Figure~\ref{fig:d4sel} we compare our D4000 cut (red region) with the rest-frame \ur\ colour selection (blue region)  and we can see that  between $0.4\lesssim z\lesssim1$ there is a distinct shallowing of $\alpha$ applying our colour cut, which is a result of the flattening of the SF-MS at stellar masses $\gtrsim10.5 \ M_{\odot}$. 

\begin{figure*}
\centering
 \includegraphics[width=1\textwidth]{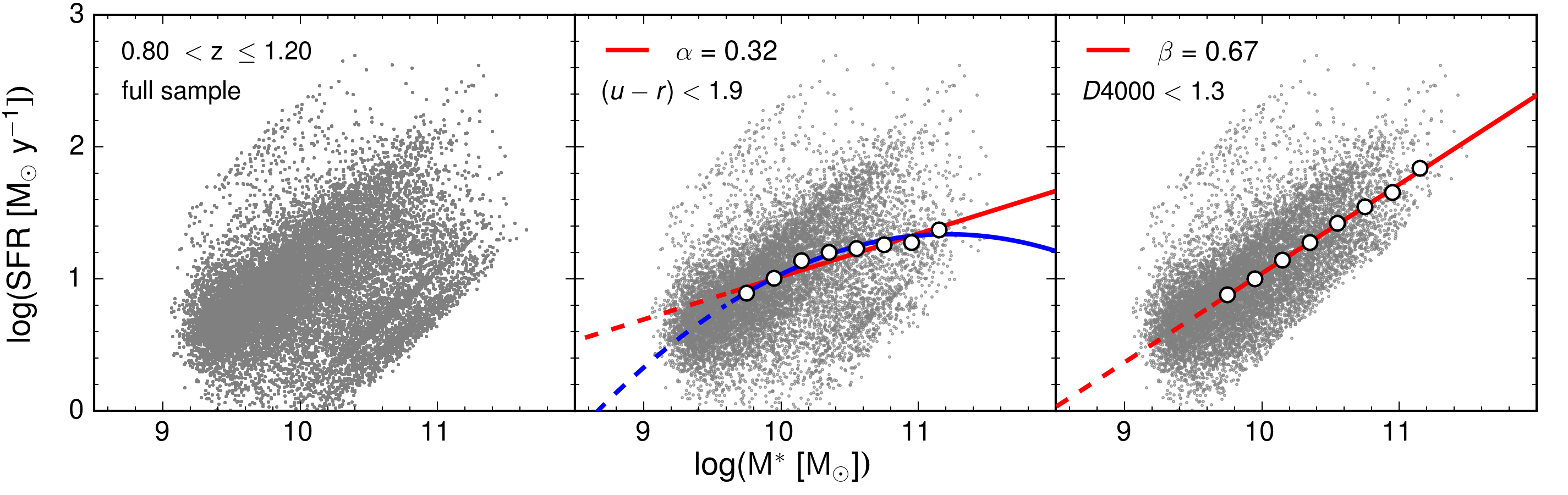}
     \caption{\small Comparison of the impact of different SFG selection methods on the SF-MS. The left panel shows the full sample at $z\sim1$ that includes a passive population of galaxies. In the middle panel we show a star-forming sample applying a rest-frame \ur\ colour cut. The white solid circles are the median points of the distribution. The red solid line shows the resulting power-law fit for the stellar mass-complete sample yielding a slope of $\alpha=0.29$; the dashed line is just the extrapolation of this fit below the completeness limit. The blue line shows a quadratic fit to the same median points which yielded coefficients of $A1,A2,A3$=[-21.527, 4.063, -0.1807] (see Equation~\ref{equ:magnelli}). The right-hand panel now shows the star-forming sample if we use a D4000 index cut. The red line here shows a power-law slope of $\alpha=0.66$.}
    \label{fig:sfcut}    
\end{figure*} 

This is shown more clearly in  Figure~\ref{fig:sfcut} where we show the SFR versus \mstar\ relation, over the redshift slice $0.8<z\leq1.2$. The left-hand panel shows the full SFR-\mstar\ distribution prior to any selection for SF galaxies. In the middle panel we show the resulting distribution once we impose the \ur\ colour cut, and finally the right panel shows our D4000 cut. It is clear that with a \ur\ colour cut there remains a residual population that, with the D4000 selection, would be considered a more passive population. This results in the median points having lower SFRs toward the high mass end. Thus, the resulting fit is shallower than that of the D4000 cut. Using the \ur\ we thus find similar model constraints to those of \cite{Magnelli2014AA...561A..86M} and \cite{Whitaker2014ApJ...795..104W} applying the quadratic model of Equation~\ref{equ:magnelli}.  This fit is shown as the blue line where we find [$A1, A2, A3$] = $[-24.596,  4.626,  -0.206]$. Over the same redshift range  \citet{Magnelli2014AA...561A..86M}  found [-26.51, 4.77, -0.202] and in  \cite{Whitaker2014ApJ...795..104W} they found [$-27.4\pm1.9, 5.0\pm0.4, -0.22\pm0.02$] at $0.5<z<1.0$ and [$-26.0\pm1.7, 4.6\pm0.3, -0.19\pm0.02$] at $1.0<z<1.5$.  
Although our results using a \ur\ colour selection show consistency in the model parameter fits to previous work, this is more likely due to residual contamination by an older sub-population of galaxies due to applying the \ur\ cut.

In addition, if we apply a more relaxed cut in D4000 from 1.3 to 1.35, thus allowing a slightly older galaxy population into our sample we can also observe the turnover in our median trends. This is demonstrated  in the right hand panel of Figure~\ref{fig:d4sel}, where the black stars are our fiducial result of D4000$<$1.3 and the red squares are from \cite{Whitaker2014ApJ...795..104W}. As we can see, with the cut at D4000 $<$1.3, our results are consistent with a constant slope out to high mass. However, with a cut at D4000 $<$1.35, the slope flattens at  \lgmstar$\gtrsim$10.5. These tests highlight our sensitivity of our fits to the initial selection of  SFGs.

Lastly, if we look at Table~\ref{tab:assumps} the majority of works we compare to (including our own) have selected samples  over an area of $\lesssim1$ deg$^2$. At these scales uncertainties  due to cosmic variance can become important particularly at the high mass end. In our work and that of e.g. \cite{Ilbert:2013AA...556A..55I} this has been included in the model constraints, however, it is unclear as to whether other works consider this effect in their analysis.

\section{Comparison with simulations}\label{sec:sims}
In this section we compare our results for the SF-MS with a variety of simulations, performed with a number of different methods, to help elucidate the key physical processes that dictate the evolution of massive galaxies. 

\subsection{Hydrodynamical simulations}
From the hydrodynamical approach we look at the Horizon \citep{Dubois2014MNRAS.444.1453D},  Illustris  \citep{Sparre:2014arXiv1409.0009S} and the  \citet{Dave2013MNRAS.434.2645D} (D13) simulations.  As well as hydrodynamics and gravity, this genre of simulations typically  include  varying prescriptions of star formation, gas cooling and heating, and feedback from stellar winds, supernovae and AGN.  Discussion on the  relative merits of the various simulations is beyond the scope of this paper, however,  we do note that they each probe varying volumes   and resolutions:
Illustris use a box size of 106.5 \mph\ on the side with 1820$^3$ DM particles and 1820$^3$ gas cells with particle and gas target masses of $6.26\times10^6\ M_\odot$ respectivley.  Horizon  have similar volume  of 100 \mph\ on the side with 1024$^3$ dark matter (DM) particles of  mass $8 \times 10^7\ M_\odot$, and  D13  used the  \textsc{Gadget}-2 N-body + Smoothed Particle Hydrodynamics \cite{Springel:2005} to evolve their simulations, within a volume of 32 \mph\ on the side containing 512$^3$  gas particles of mass $4.5\times10^6\ M_\odot$.

\subsection{Scaling relation approaches}
\citet{Behroozi2013ApJ...770...57B} parameterised the stellar mass--halo mass (\mstar-$M_h$; SMHM) scaling relation to model observed star-formation histories out to $z=8$. By  applying a MCMC approach to constrain their 15 free parameters, they begin with an initial guess for the SMHM relation, then find galaxy growth histories by applying the SMHM relation to dark matter merger trees. They then derive a series of inferred stellar mass functions and SFRs onto which they apply simulated observational errors and biases. Finally they compare their model to available data to constrain the SMHM relation.

The recent work by  \citet{Mitra:2014arXiv1411.1157M} (M14) is an analytical based approach which also exploits the use of observed  scaling relations. In their approach they begin with a mass balance equation to model the inflows and outflows in the interstellar medium (ISM) assuming galaxies grow along a slowly-evolving equilibrium between accretion, feedback  and star formation \citep[e.g.][]{Finlator2008MNRAS.385.2181F}. In essence, the parameters of the equilibrium model describe the motion of gas into and out of galaxies, which is termed  the {\it baryon cycle}.  In their modelling they have  only 8 free parameters that are constrained via a MCMC,  comparing their model to the   \mstar-$M_h$ constraints of   \cite{Behroozi2013ApJ...770...57B} and \cite{Moster2013MNRAS.428.3121M},   as well as the  \mstar-metallicity scaling relation from e.g. \citet{Andrews2013ApJ...765..140A,Zahid2014ApJ...791..130Z} and \cite{Steidel2014ApJ...795..165S}, out to $z\sim2$.

Thus, by design, these types of approaches should be able to better reproduce the observed SFR-\mstar\ relations compared to hydrodynamical and SAM simulations.

\subsection{Results}
\begin{figure*}
\centering
 \includegraphics[width=0.98\textwidth]{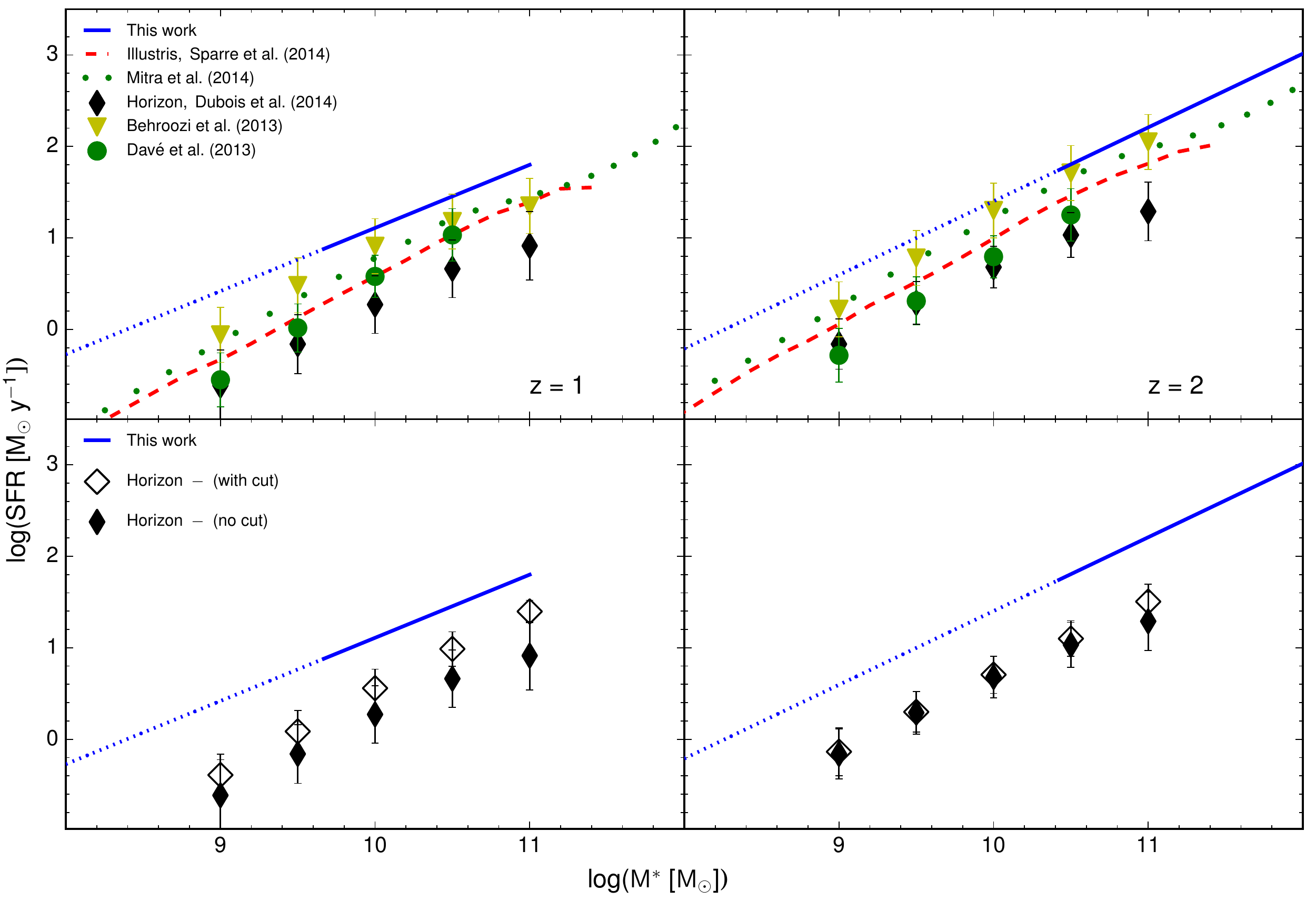}
     \caption{\small  Comparison of the SFR-MS with current simulations.  The shaded distribution of points with contours show our SFR-Mass results at $z\sim 1$ (left) and $z\sim 2$ (right). The fit to our data is shown as the solid blue line, with the dashed line extrapolated below our stellar mass completeness limit. The results from simulations at these redshifts are Illustris \citep[][red solid line, hydro]{Sparre:2014arXiv1409.0009S},      \citet{Mitra:2014arXiv1411.1157M} (green solid line, equilibrium model),  Horizon   \citep[][black diamonds, hydro]{Dubois2014MNRAS.444.1453D}, \citet{Dave2013MNRAS.434.2645D} (green dots, hydro) and \citet{Behroozi2013ApJ...770...57B} (yellow triangles). }
    \label{fig:sims}    
\end{figure*}

In Figure~\ref{fig:sims} we show the SF-MS at $z=1$ (left) and 2  (right). The blue solid line shows the fit to our data above our stellar mass completeness limit. The dotted blue line is an extrapolation of this fit for illustration purposes. The simulations show general agreement with the slope of the relation, but all have a lower normalisation in SFR by varying degrees. Perhaps as expected, the scaling relation approaches are most aligned with our work. By averaging over the stellar mass bins at 10, 10.5 and 11.0 we find M14 and B13 to be lower by a factor $\sim1.8$. In contrast we find that the SF-MS from the hydrodynamics simulations of D13 and Illustris to be lower by as much as $\sim 3$ with Horizon showing the largest discrepancy being lower by a factor of $\sim$6.  At $z=2$ there is  improvement in the normalisation of M14 and B13, with their models now being lower by factors of 1.4 and 1.2 respectively between \mstar\ of 10.5 and 11.  The hydrodynamics predictions show only a slight change from $z=1$, and continue to significant under-predict the normalisation in the SF-MS relation.
  
It is worth noting that the Horizon simulation data set available to us, contains both SFGs and passive galaxies.  Although we are unable to segregate them explicitly, we artificially impose a cut in SFR that closely mirrors our D4000 selection, shown in the bottom panels of Figure~\ref{fig:sims}. As we can see there is only a marginal upward shift in the normalisation at $z=1$ and very little change at $z=2$, implying  that the bulk of the simulated galaxies have significantly lower SFRs at a given stellar mass compared to the observations.

\begin{figure}
\centering
 \includegraphics[width=0.45\textwidth]{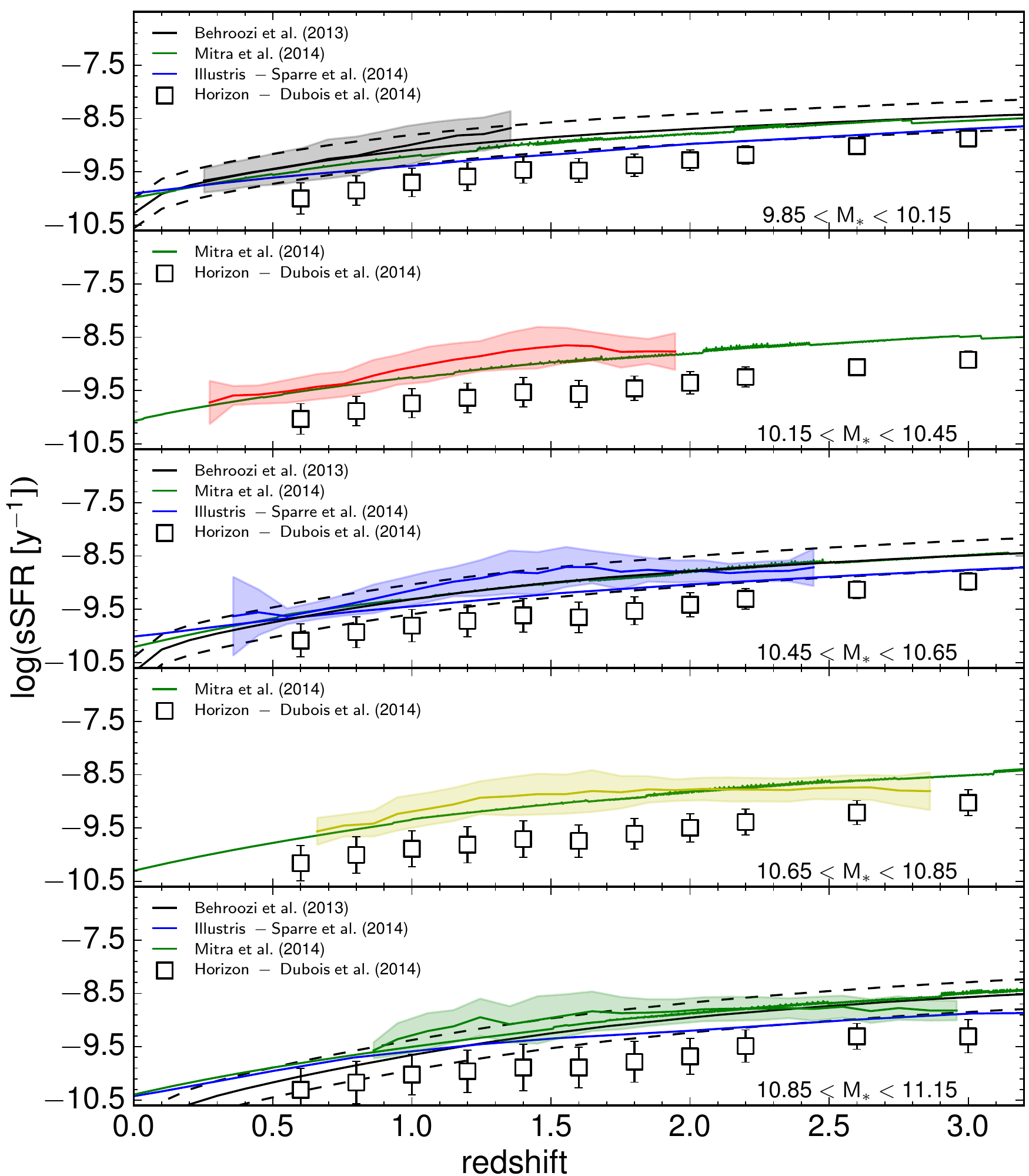}
     \caption{\small   Same as Figure~\ref{fig:ssfr-z} but now comparing  to simulations.  Predictions  are from  \citet{Behroozi2013ApJ...770...57B}, \citet{Mitra:2014arXiv1411.1157M},  \citep[Illustris,][]{Genel2014MNRAS.445..175G,Sparre:2014arXiv1409.0009S}  and  \citep[Horizon,][]{Dubois2014MNRAS.444.1453D}
     }\label{fig:ssfr-sims}
\end{figure}

We now compare the observational results on the  evolution of the sSFR (Figure~\ref{fig:ssfr-sims}),  to the simulations, for the same  range of stellar mass bins considered in \S~\ref{sec:ssfr}.  If we consider redshifts $z<1.0$ across all stellar mass bins, we find that M14, B13 and Illustris show good agreement with our results. At $1<z<2$ the Illustris predictions show an average factor of $\sim3$ lower normalisation at \lgmstar=10, which seems to reduce to 2 towards the highest stellar mass bin \mstar= 11.  We find both M14 and B13 continue to show very good agreement with our results across all stellar mass bins at $z>1$.   Finally, we again note that the Horizon simulations are lower by a factor of $\sim6$ across all mass bins.

\section{Discussion} \label{sec:discussion}

Theoretical predictions from the $\Lambda$CDM   modelling of the average growth rate of dark matter halos through smooth cold accretion was well  characterised to be  $\propto(1+z)^{2.25}$   at $z\sim2$ \citep{2009Natur.457..451D}.  Developments in this area now favour a stronger evolution in sSFR to $\propto(1+z)^{2.5}$ \citep[][]{Dekel2013MNRAS.435..999D,Faucher2011MNRAS.417.2982F} which while lower than our constraints, is beginning to show convergence toward a higher evolution.
 In our comparison  of the  SFR-\mstar\ and sSFR-z trends with the hydrodynamical  simulations of  Illustris  \citep{Sparre:2014arXiv1409.0009S}, Horizon \citep{Dubois2014MNRAS.444.1453D} and \cite{Dave2013MNRAS.434.2645D} we found    in all cases an  under prediction of the normalisation  by a factor ranging between $\sim2$ \citep[llustris and][]{Dave2013MNRAS.434.2645D} and  6 (Horizon).  Such a disparity between observations and simulations, both in hydrodynamical and  semi analytical modelling (SAMs), has  been identified and discussed extensively \citep[e.g.][]{daddi07,Elbaz2007AA...468...33E,Santini2009AA...504..751S,Damen2009ApJ...705..617D,Dave2013MNRAS.434.2645D,Sparre:2014arXiv1409.0009S,Genel2014MNRAS.445..175G,Tasca2014arXiv1411.5687T}.  The general consensus  shows at low redshifts ($z\lesssim0.5$) simulations can reproduce the sSFR and again at $z\gtrsim4$. However in the intermediate regime there remains a disconnect in the evolution of sSFR. Whilst there are a number of possible issues that may account for this effect  that have been suggested such as: oversimplified   gas accretion modelling,  systematic offsets in gas cooling rates, insufficient sub-grid models that control star formation and stellar feedback, this remains an unresolved issue. Moreover, hydrodynamical and SAMs typically require $\sim20$ and  $\sim30$-50 parameters  respectively which can make drawing physical conclusions challenging.

Fundamentally, observations seem to be increasingly  favouring the downsizing scenario, whereas  simulated star formation histories are tied to the  dark matter accretion histories and cannot produce the observed mass-dependence in sSFR \citep{Damen2009ApJ...705..617D,Somerville2014arXiv1412.2712S,Sparre:2014arXiv1409.0009S}. 

The scaling relation based approaches of \citet{Behroozi2013ApJ...770...57B} and  \cite{Mitra:2014arXiv1411.1157M} proved they could recover well the sSFR evolution across all mass ranges. This is perhaps not too surprising since their models are being constrained by current observational data. Nevertheless, they do present the opportunity to explore key physical aspects of galaxy formation within a relatively simple analytical framework compared to the full blown architecture of hydro and SAMs. In particular, the equilibrium model of  \cite{Mitra:2014arXiv1411.1157M} does not explicitly model halos, cooling, mergers or a disk star formation law, but instead parameterises  the motion of gas into and out of galaxies,  assuming that the gas reservoir in a galaxy is slowly evolving with an equilibrium between accretion, feedback, and star formation \citep{Finlator2008MNRAS.385.2181F,Bouche2010ApJ...718.1001B}. Their results suggest that mergers are sub-dominant for overall galaxy growth  and that the key drivers for the average evolution of SFGs is the continual smooth accretion regulated  by continual outflows with occasional cycling.

Of course the way in which we select our SFGs plays a crucial role in the derived slopes of the SF-MS.   
In this work we demonstrated how two different SFG selection criteria can lead to very different constraints in the modelled SF-MS and, in particular, the high-mass turnover. From our comparisons to a wide range of related works there is a clear and well known  bi-modal distribution in the slope of the SF-MS particularly beyond $z\sim1$; where some find $\alpha$ around 0.5 and others close to unity. In \cite{Whitaker2014ApJ...795..104W} using UV+IR derived SFRs they found a non evolving low mass (\lgmstar$<10.2$) MS slope close to unity. In contrast, they found an evolving and shallower high mass (log[\mstar/$M_{\odot}]>$10.2) slope that evolved to $\alpha=0.59$ at $z\sim2.25$. A similar turn-over has also recently been reported by \cite{Heinis:2014MNRAS.437.1268H,Magnelli2014AA...561A..86M,Schreiber2015AA...575A..74S}. In  \cite{Tasca2014arXiv1411.5687T} this has been interpreted  as evidence of a more gradual star formation quenching progressing from high masses at high redshifts toward low masses at low redshifts. In the recent work by \cite{Buat:2014AA...561A..39B} showed that the type of SFR indicator affects the normalisation of the MS and not the slope. However as discussed in e.g. \cite{Speagle2014ApJS..214...15S,Ilbert:2013AA...556A..55I} they way in which we select SFG can impact the slope considerably.

Our main approach was to select SFGs based on the D4000 index as outputted by \cig. This selection imposed a rather sharp cut in the SFR-\mstar\ distributions which  led to   a roughly  constant slope being derived   out to $z\sim1.7$ and little evidence for a high mass-turn over. However, when we instead applied a rest-frame \ur\ colour cut to isolate SFGs, we found strong evidence for a turn over at \lgmstar$\sim10.5$ which was consistent with the trends of the aforementioned works.  However, we note that in both scenarios we found slopes of $\alpha\sim0.8$ beyond $z\gtrsim2$  that showed no sign of flattening at high mass which seems more inline with the trends of \cite{Schreiber2015AA...575A..74S}.

 The issue  of which colour selection one should  adopt  to obtain a {\it clean} SF-MS has also  been highlighted in e.g. \cite{Ilbert:2013AA...556A..55I} and \cite{Speagle2014ApJS..214...15S}. Whilst using a mixture of different colour cuts to remove purely quiescent galaxies  as in e.g. \cite{Ilbert:2013AA...556A..55I} ({\it r-J, {\rm NUV}-r}) does seem to provide a cleaner selection, the use of a D4000 cut within the SED fitting framework potentially gives us a more {\it physical} way of identifying an ageing population.  This has been explored in recent work by \cite{Morokuma2015arXiv150102915M} where they found  the D4000 break as a cleaner way to select SFGs, albeit at low redshifts $z<0.2$.  Clearly this remains an ongoing issue that requires further investigation. 

\section{Conclusions}
In this study we have used CFHTLS visible $ugriz$ and VIDEO near-infrared $Z$, $Y$, $J$, $H$, $K_{\rm s}$ photometry, along with the IRAC 3.6 and 4.5 \mic\ bands from SERVS to investigate the evolution in the relation between stellar mass and star-formation rate to $z\sim 3$ using the public SED fitting code \cig\ to determine the SFR and stellar mass.  Where available we included observations from SWIRE and HerMES for the longer wavelength data. \cig\ enforces an energetic balance between dust-enshrouded UV stellar emission and its re-emission in the IR, making the most of our multi-wavelength dataset. 

We summarise our findings as follows:
\begin{itemize}

\item We propagated the full photometric redshift uncertainties in the form of their PDFs through \cig\ to provide more robust measurement of the SFR and stellar masses. In this work, the PDFs were estimated using \lephare\ and in general our results showed  good consistency with an analysis which does not consider the PDFs, with no evidence for residual bias.  

\item By modelling the data with a simple power law we have found strong evolution in the normalisation of the SFR-\mstar\ relation out to $z\sim 3$. For a fixed stellar mass of \lgmstar=10.0 we find the median SFR decreases by a factor of $\sim19$ from $z\sim2$ to $z\sim0.2$. This is consistent with the picture of quenching of star formation and downsizing. 

\item By selecting SFGs using a D4000 index cut we observe  a nearly constant slope for the SF-MS out to $z\sim1.7$ of $\alpha=0.69\pm0.02$, beyond which we observe a rise, with $\alpha = 0.90\pm0.03$ at $z\sim2.2$.  At the highest redshift range probed, $2.3<z<3.0$ we find  slightly lower values for the slope of $\alpha=0.80\pm0.03$.   

\item However, when selecting on the basis of rest-frame \ur\ colour rather than the D4000 index,  we found  a slope that became progressively shallower out to $z\lesssim1.7$, reaching a low value of $\beta\sim0.35$ at $z\sim1$.  We found that in this scenario the SFR-\mstar\ relation was more appropriately modelled by a broken power law or a quadratic, showing a turnover at \lgmstar$\sim$10.5. This effect was also observed  by relaxing our D4000 cut from 1.3 to 1.35 and thus allowing a slightly older population into our SF sample. On closer inspection we found that the \ur\ selection is not as efficient at removing  the passive  population and the turnover is largely due to a residual quiescent population that is not isolated. However, beyond $z\gtrsim1.7$ our results are consistent with the trends of the D4000$<$1.3 selection,  showing a steepening of $\alpha\sim0.8$ out to $z\lesssim3$.

\item Our study of the sSFR with redshift  from $z=0.1$ to 3 found strong evolution.  By modelling the data as  sSFR $\propto(1+z)^\gamma$ we find a similar mass-dependent evolution out to $z\lesssim1.4$  with  $\gamma=3.39\pm0.01$ to $\gamma=3.91\pm0.36$ from \lgmstar=9.75 to 11.25 respectively. Extending this analysis to higher redshifts yielded  $\gamma=2.60\pm0.04$ for a stellar fixed  mass of \lgmstar=10.5 out to $z\sim2.44$.  At our highest stellar mass bin of \lgmstar=11 we found a flattening in the relation  with $\gamma=2.13\pm0.06$ between $0.86<z\lesssim3$.

\end{itemize}

Finally, our current sample size has been limited to 1 deg$^2$ to allow  comprehensive broadband coverage.  However, the VIDEO survey will have a total coverage of 12 deg$^2$ on its completion, allowing us to probe a larger and deeper  area and thus will provide some of the most concise measurements of the SFR-\mstar\ relation, leading to greater insight into how  this relates to   the underlying dark matter distribution and its evolution  with redshift.

\section*{Acknowledgments}
{\small 
We would like to thank Veronique Buat, Denis Burgarella,  Romeel Dav\' e, Claudia Maraston, David Gilbank, Ros Skelton, Martin Hendry,  Julien Devriendt, Duncan Farrah, Dan Smith, Daniel Cunnama and Jonny Zwart for useful discussions throughout this project. We are grateful to  Giulia Rodighiero, Catherine Whitaker, Olivier Ilbert, Peter Behroozi, Sebastien Heinis, the Horizon and Illustris simulation teams,  
Sourav Mitra and Romeel Dav\' e for providing their data products which proved to most helpful in our analysis.
The bulk of observational sSFR data from Figure~\ref{fig:ssfr-z}  were obtained from \url{http://www.peterbehroozi.com/data.html}}

{\small
RJ, MV, MJ, MP, MS and EG acknowledge the support from the South African National Research Foundation and the Square Kilometre Array (South Africa). MV and MJ acknowledge support from the European Commission Research Executive Agency FP7-SPACE-2013-1 Scheme (Grant Agreement 607254)}

{\small
Numerical computations were done on the Sciama High 
Performance Compute (HPC) cluster which is supported by the ICG, SEPNet and the University of
Portsmouth. }

{\small
Based on data products from observations made with ESO Telescopes at the La Silla or Paranal Observatories under ESO programme ID 179.A-2006 and was supported by the European Commission Research Executive Agency FP7-SPACE-2013-1 Scheme (Grant Agreement 607254 - Herschel Extragalactic Legacy Project - HELP).
%

%
%
%
\setcounter{equation}{0}
\renewcommand{\theequation}{A-\arabic{equation}}
\setcounter{section}{0}
\renewcommand{\thesection}{A-\arabic{section}}
\setcounter{figure}{0}
\renewcommand{\thefigure}{A-\arabic{figure}}
\bibliographystyle{mnras}

\bibliography{bibDB}

\bsp	
\label{lastpage}
\end{document}